\documentclass{aastex63}

\usepackage{xcolor}

\received{}
\revised{}
\accepted{}

\submitjournal{ApJ}

\shorttitle{Machine-learning approach to identification of coronal holes}
\shortauthors{Illarionov et al.}

\graphicspath{{./}{figures/}}

\begin{document}

\title{Machine-learning approach to identification of coronal holes
in solar disk images and synoptic maps}

\correspondingauthor{Egor Illarionov}
\email{egor.mypost@gmail.com}

\author{Egor Illarionov}
\affiliation{Moscow State University, Moscow, 119991, Russia}
\affiliation{Moscow Center of Fundamental and Applied Mathematics, Moscow, 119234, Russia}

\author{Alexander Kosovichev}
\affiliation{Center for Computational Heliophysics, New Jersey Institute of Technology, Newark, NJ 07102, USA}
\affiliation{Department of Physics, New Jersey Institute of Technology, Newark, NJ 07102, USA}
\affiliation{NASA Ames Research Center, Moffett Field, CA 94035, USA}

\author{Andrey Tlatov}
\affiliation{Kislovodsk Mountain Astronomical Station of the Pulkovo Observatory, Kislovodsk, 357700, Russia}

\begin{abstract}

Identification of solar coronal holes (CHs) provides information
both for operational space weather forecasting and long-term
investigation of solar activity. Source data for the first
problem are typically most recent solar disk observations,
while for the second problem it is convenient to consider solar
synoptic maps.
Motivated by the idea that
the concept of CHs should be similar for both cases
we investigate universal models that can
learn a CHs segmentation in disk images and reproduce the same 
segmentation in synoptic maps. We demonstrate
that Convolutional Neural Networks (CNN) trained on daily disk images 
provide an accurate CHs segmentation in synoptic maps and 
their pole-centric projections.
Using this approach we construct a catalog
of synoptic maps for the period of 2010--20 based on SDO/AIA observations in the 193~Angstrom wavelength. The obtained CHs synoptic maps are compared with magnetic
synoptic maps in the time-latitude and time-longitude diagrams.
The initial results demonstrate that while in some cases the CHs are associated with magnetic
flux transport events there are other mechanisms contributing to the CHs formation
and evolution.  
To stimulate further investigations the catalog of synoptic maps is published in open access.

\end{abstract}


\section{Introduction} \label{sec:intro}

Solar magnetic fields play a key role in the formation of
solar activity tracers that are observed in solar disk
images \citep{Solanki_2006}. Regions, where magnetic field
lines are open in the outer space and appear darker in EUV images,
are called coronal holes (CHs).
Direct observation of such structures is a 
challenging procedure and requires special conditions
\citep{Lin_2004}. Another option based on a reconstruction
of magnetic field lines from solar magnetograms requires
additional modeling \citep[see e.g.][for details of observations]{Stenflo_2013}.
There are long-term and intense debates about a proper
way of the magnetic field reconstruction and there is no single accepted way
\citep[see][for review of models and its limitations]{Wiegelmann_2014, Wiegelmann_2017}.

A search for a robust detection procedure for CHs is
motivated by at least two aspects. First, due to the open magnetic
field line configuration, high-energy particles can easily
flow into the outer space and form a solar wind
\citep{Nolte1976, Abramenko2009, Cranmer_2009, Obridko2009}.
The solar wind from  CHs can reach the 
Earth and manifest itself in geomagnetic storms \citep{Robbins2006, Vrsnak_2007}. 
Thus, the detection of CHs is essential for space weather
forecasting. Second, in the view of the solar dynamo theory,
periods of solar activity minima are associated with a strong poloidal
magnetic field \citep{Parker_1955}.
Thus, observations of polar CHs may provide information
about the poloidal field strength and also the upcoming 
solar cycle \citep{Harvey_2002}. Identification of CHs as open
field regions in reconstructed solar magnetic field lines is
doable, however, with significant uncertainties \citep[see e.g.][]{Linker_2017}

Fortunately, CHs have an easily accessible tracer.
They appear as massive dark regions when the solar disk
is observed in the EUV or X-ray spectrum. The reason for
its darker appearance is a lower density and temperature
of the solar corona due to the special magnetic field configuration \citep{priest2014}.
Detection of such specific dark regions is a convenient way for
CH identification. We review some common approaches to this
problem below. 

Detection of CHs is performed both in solar disk images and
in solar synoptic (Carrington) maps that are a compilation of
successive disk images during a solar rotation period.
Methods for CHs identification in the disk images are remarkably diverse.
They range from fully manual procedures to fully automatic ones
and use observations in various wavelengths (Table \ref{tab:ch_methods}). 
In addition, source data providers often apply a custom data preprocessing
that contributes to disagreements among various identification attempts. 
A detailed and unbiased analysis of the various approaches and their 
uncertainties is outside the scope of this research.

\begin{deluxetable*}{cccc}
\tablenum{1}
\tablecaption{Input data used for CHs segmentation in previous studies.\label{tab:ch_methods}}
\tablehead{
\colhead{Author} & \colhead{Reference name} & \colhead{Input wavelength}
}
\startdata
\citet{Henney_2005} & -- & 10830~{\AA} and magnetogram \\
\citet{Scholl2008} & -- & 171, 195, 304~{\AA} and magnetogram \\
\citet{Krista2009} & -- & 195~{\AA} \\
\citet{Reiss_2014} & -- & 193~{\AA} \\
\citet{Verbeeck_2014} & SPoCA & 193~{\AA} or 195~{\AA} or (171 and 195~{\AA}) \\
\citet{Lowder_2017} & -- & (193 or 195~{\AA}) and magnetogram \\
\citet{Garton_2018} & CHIMERA & 171, 193 and 211~{\AA} \\
\citet{Heinemann_2019} & CATCH &  193~{\AA} and magnetogram
\enddata
\end{deluxetable*}

Further progress in methods for CHs identification
in disk images will help to reduce uncertainties in the determination 
of CH boundaries.
However, CHs are typically large structures, and a single disk image may
reveal only a part of a CH that is on the visible side of the Sun.
This means that we need some compilation of series of disk images to
capture the whole region of CH. Solar synoptic maps are a convenient
way for such representation. A straightforward approach to get the CHs
boundaries in a synoptic map is a compilation of the CHs boundaries
detected in disk images. This approach was implemented e.g. by 
\citet{Caplan_2016}. We note that this approach may unambiguously work 
only if all disk images are taken at the same time and cover the
whole solar surface. However, CHs evolve and change their shape with time.
Even long-living CHs may appear substantially different
in the disk images after a single solar rotation. 
The instantaneous coverage of the whole solar surface was only 
available during the STEREO observations of the far-side of the Sun.

An alternative approach suggests first to merge the solar disk images
into full-surface synoptic maps, and then identify CHs in the synoptic
map directly. Of course, we still have uncertainties
in pixel intensities, however, it is more convenient to resolve
them for continuous values (pixel intensities) than for
binary values (CH boundaries). Quite surprisingly we find
much less recent publications on CH identification in the synoptic
maps. \citet{Toma_2005} and \citet{Toma_2011} developed a CH identification
procedure using synoptic maps in the 171, 195, 284, 304, 10830~{\AA}
spectrum lines along with the H$\alpha$ and magnetic synoptic maps.
The dataset and analysis cover a period from 2006 to 2009.
\citet{Webber_2014} investigated polar coronal holes from 1996
through 2010 and compared the identification of CHs in the disk images
with two techniques that identify CHs in the synoptic maps.
One method is based on a combination of synoptic maps in the 171, 195,
and 304~{\AA} wavelengths, while the second one works with the magnetic
synoptic maps. The authors concluded that these methods
produced comparable results. 
An extended time-period from 1996 to
2016 was considered by \citet{Hamada_2018} who used
the multi-wavelength synoptic maps  together with magnetograms. 
An important contribution of this paper is the development of a homogenization procedure 
for data from different observational instruments, which allowed them to perform
a joint analysis of two solar cycles (23 and 24).

The previous methods were developed to analyze specifically either
disk images or synoptic maps.  We did not find any method that 
has been validated both in solar disk images and synoptic maps.
This motivates us to develop a unified procedure that can
be applied to various representations of solar observations.

In this paper, we suggest an idea that for a unified detection algorithm
there should be no dramatic difference between CHs captured
in solar disk images and synoptic maps. Of course, we
appreciate that the CHs in the disk images are physical objects while
in the synoptic maps they are synthetic objects to some extent. 
Nevertheless, visual interpretation works similarly in both cases.
One can say that the concept of CHs is the same in both representations.

The suggested idea provides some desired properties of the
unified algorithm. First, it should be local in the sense
that it should be independent on the global image scale
and context. For example, it should demonstrate the same output
whether we feed a whole solar disk or just a cropped patch with
no information about its location in the original disk image.
Second, reasonable geometrical transformations should not affect the
CH identification, e.g. there should be no difference to which plane the
solar sphere is projected, -- the concept of CH remains the same.

Analyzing the desired properties we note that if the algorithm
acts as a convolution with some local kernel, it can be a proper
candidate. Of course, the kernel should be sophisticated enough
to provide binary masks of CHs from input images. This
is very close to what Convolutional Neural Networks (CNN) do.

The CNN are a special type of neural networks commonly used in 
image analysis. They can be assumed as a set of successive
convolutional operations with the kernels that are adjusted during
a model training phase to minimize some loss function,
e.g. a segmentation error. Once the model is trained,
the kernels are fixed and inference in new images can be done.
A nice and useful property of such models is that due to their
architecture they do not depend on the input image size (the situation
is similar to the well-known Gaussian or Sobel filters that
can be applied to images of arbitrary shape).

In our research, we apply a CNN trained on segmentation of CHs
in solar disk images to solar synoptic maps. We present an 
algorithm of solar synoptic maps construction and demonstrate
that the CNN model provides an accurate segmentation output.
As a special case, we consider synoptic maps projected onto the Northern
and Southern solar hemispheres (pole-centric projections) and
demonstrate that the output of the CNN model is also in
agreement with the original synoptic map. The obtained statistics
is analyzed with respect to solar activity variations.

\section{Data} \label{sec:data}

We analyze a dataset of the Solar Dynamics Observatory (SDO)
Atmospheric Imaging Assembly (AIA) 193~{\AA} solar disk images
with a cadence of one image per day \citep{Lemen2012}.
Start date is 2010-06-16, the end date is 2020-03-01.
This period covers 130 full solar rotation periods starting from
Carrington rotation (CR) number 2098 to 2227 inclusively.
The dataset was obtained from the SunInTime\footnote{\url{https://suntoday.lmsal.com/suntoday/}}
website in JPEG quality and 1K resolution. There are two reasons
for this choice. First, this is the same dataset
as was used by \citet{Illarionov_2018} for the CNN model training.
In the context of neural networks models, the dataset uniformity
is essential. Second, this dataset is already calibrated with
respect to any known instrument issues by the instrument team \citep{Lemen2012}.
This allows a direct assessment of the input data quality and prevents
from possible misinterpretation in data preprocessing steps.
Based on this data we construct solar synoptic maps
as described in the next section.

In the data analysis section, we use also  
Carrington rotation synoptic charts of radial magnetic field
component\footnote{\url{http://hmi.stanford.edu/data/synoptic.html}} from Helioseismic
and Magnetic Imager \citep[HMI,][]{Scherrer_2012}.

Additionally, we use a catalogue of filaments\footnote{\url{https://observethesun.com/}} provided by the Kislovodsk Mountain Astronomical Station\footnote{\url{http://en.solarstation.ru/}} to make a comparison 
with CHs identified by the CNN model.

\section{Construction of synoptic maps} \label{sec:maps}
A standard way of the synoptic map construction consists of two steps.
First, we project the solar disk images onto the Carrington
coordinate system. Second, we select latitudinal strips
centered at the central meridian and concatenate 
them within a single solar rotation period. Other catalogs of 
the SDO/AIA synoptic maps were prepared similarly
\citep[e.g.][]{Karna_2014,Caplan_2016,Hamada_2020}.

For the construction of the synoptic maps, we use a dataset of
solar disk images described in Section \ref{sec:data}.
The disk images have a resolution of $1024\times1024$ pixels; the
synoptic maps are calculated with the resolution of $720\times360$
(however, this is a free parameter).
First, we map each disk image into the Carrington coordinate system.
A technical problem here is how to map  pixels of disk images 
onto synoptic
maps. On one side, for each pixel in a disk image,
one can find a corresponding pixel in the synoptic map using basic
trigonometric formulas. The advantage is that we use information
from all pixels that cover the solar disk; the disadvantage is that
the corresponding pixels of the synoptic map are sparse. The higher the
resolution of the synoptic map, the greater its sparsity.
On the other side, one can construct a reverse mapping.
The advantage here is that pixels of the synoptic map are dense,
however, some pixels of disk image will be ignored
and not contribute to the synoptic map. In this case, the higher 
the resolution of the disk image, the greater the number of pixels
ignored in this image.
Since we want to keep the resolution of the synoptic maps as a free parameter,
we suggest using the mapping of both types and averaging the pixel
values that correspond to the same pixel of a synoptic map.

The next step is to select a strip around the central meridian of
the projected disk image. It is convenient to consider this step
as a part of an averaging procedure, in which we take into 
account the distance between the pixel longitudes and 
the central meridian longitude in the contributing disk image. 
The greater the distance, the smaller the pixel weighting factor.
The proposed weighting function is defined as: ${\rm sigmoid}((-d + a) / b)$,
where ${\rm sigmoid}(x) = 1/(1+\exp(-x))$ is a standard sigmoid function,
$d$ is a distance in degrees, $a$ and $b$ are the shift and scale parameters
that help to select the desired blending.
Indeed, varying these parameters we will obtain wider or narrower
rectangular domains and can play with the softness of its borders. 
As a particular choice in this work, we use the weighting
function: ${\rm sigmoid}((-d + 13.2) / 2)$. It approximately
specifies that each pixel in the synoptic map is mostly a result of the
blending of two nearest disk images.
In Section \ref{sec:results} we will demonstrate that the CH detection is 
stable against the various choice of these parameters.
This particular choice was motivated mostly by the
visual appearance of the produced synoptic maps.
Larger values of the shift parameter result in losing fine
structures in the synoptic maps, while smaller values make the transition
zones between the successive disk images visible.

\begin{figure}
\centering
\includegraphics[width=\textwidth]{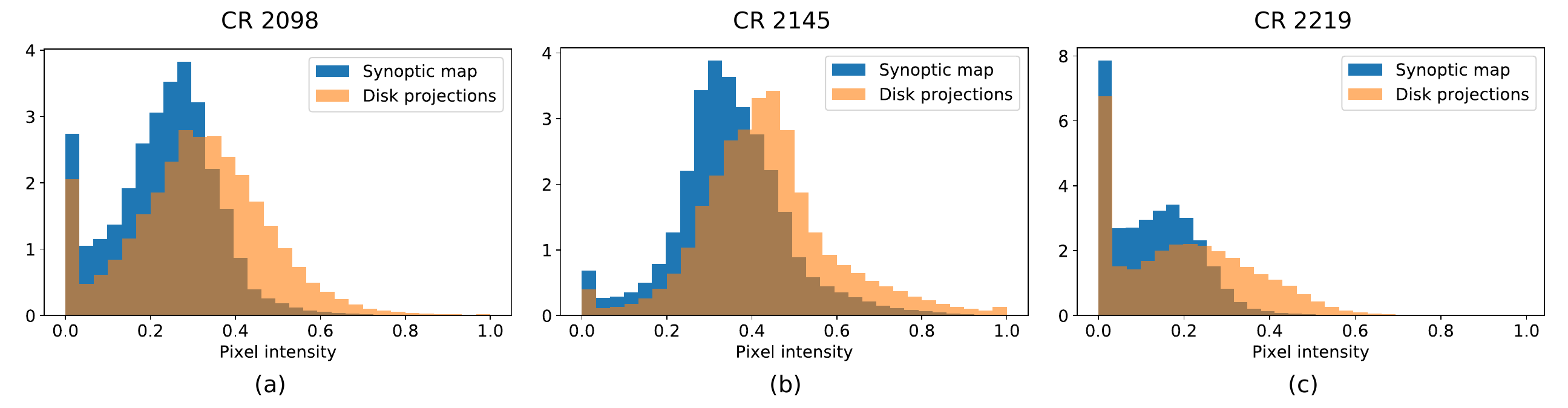}
\caption{Pixel intensity distribution of synoptic maps before histogram matching in
comparison to the distribution of contributing disk projections.
Histogram matching procedure adjusts the synoptic map to make
it similar to disk projections. Carrington rotations: a) CR~2098,
b) CR~2145 and c) CR~2219 are shown.
\label{fig:hist}}
\end{figure}

The final step is a histogram matching that corrects the
brightness and contrast variations in the disk area
due to the limb-brightening effect. Because of this effect, a synoptic map
constructed from the central meridian strips appears darker 
than the original disk images. As a result, the pixel intensity distribution
is biased. A variety of physics- and data-driven models
have been proposed for correction of this effect \citep[see, e.g.][]{Caplan_2016}.
We apply the most straightforward approach of direct histogram matching. 
First, we construct a cumulative
distribution function (CDF) $F_1$ for pixel intensities from all 
contributing projected disk images. Then we construct a CDF $F_2$
for the synoptic map. It remains to note that
if we replace each pixel intensity level $p$ of a synoptic map with
$F_1^{-1}(F_2(p))$ then we obtain a new distribution with CDF equal to $F_1$ \citep[see, e.g.][for implementation details]{Gonzalez_2006}.
Figure~\ref{fig:hist} shows the pixel intensity distributions
for a sample of synoptic maps before the histogram matching and the 
distribution of disk projections.

\begin{figure*}
\centering\includegraphics[width=0.8\textwidth]{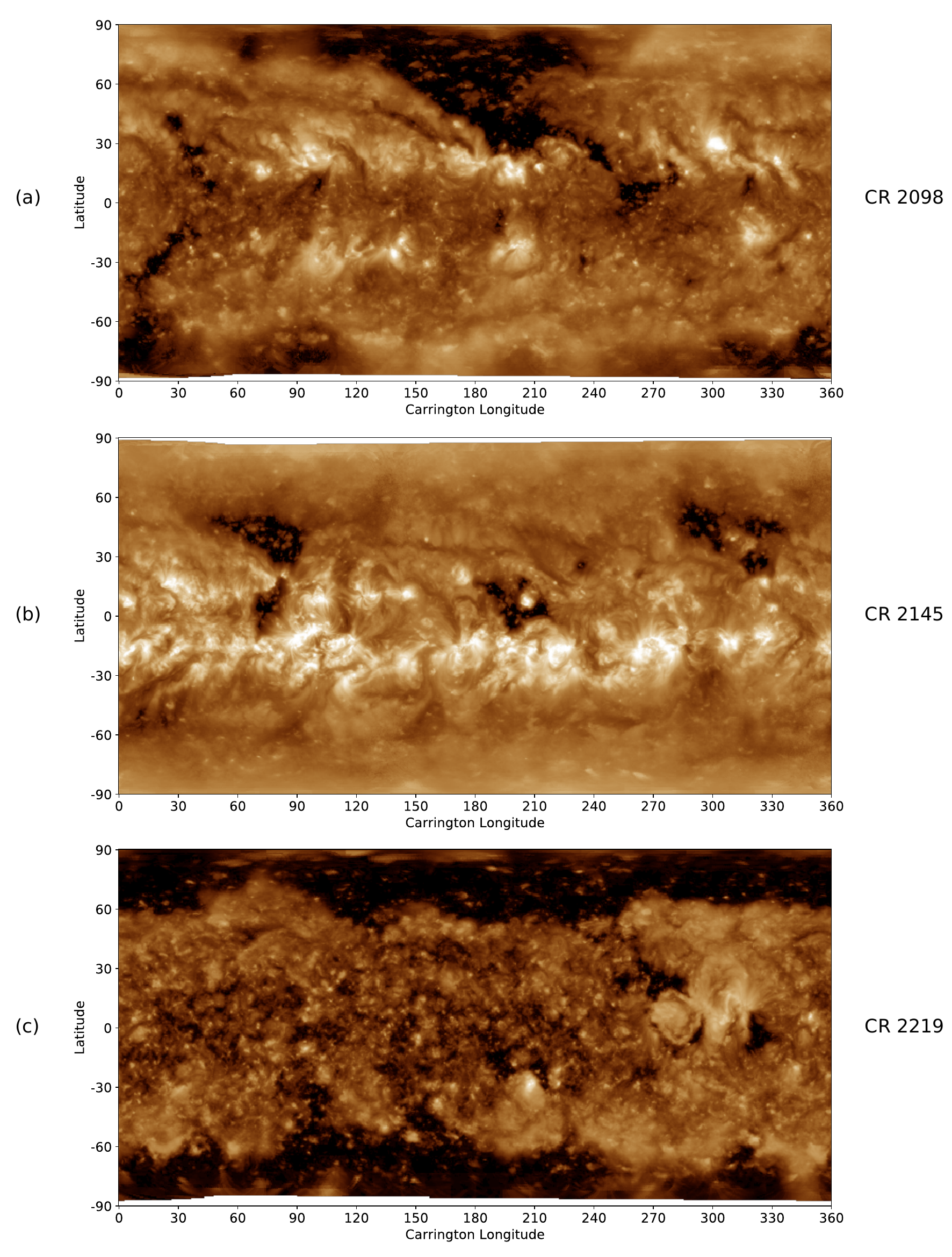}
\caption{Sample synoptic maps for Carrington rotations: a) CR~2098,  b) CR~2145, and c) CR~2219. 
\label{fig:maps}}
\end{figure*}

\begin{figure}[h!]
\centering
\includegraphics[width=0.75\textwidth]{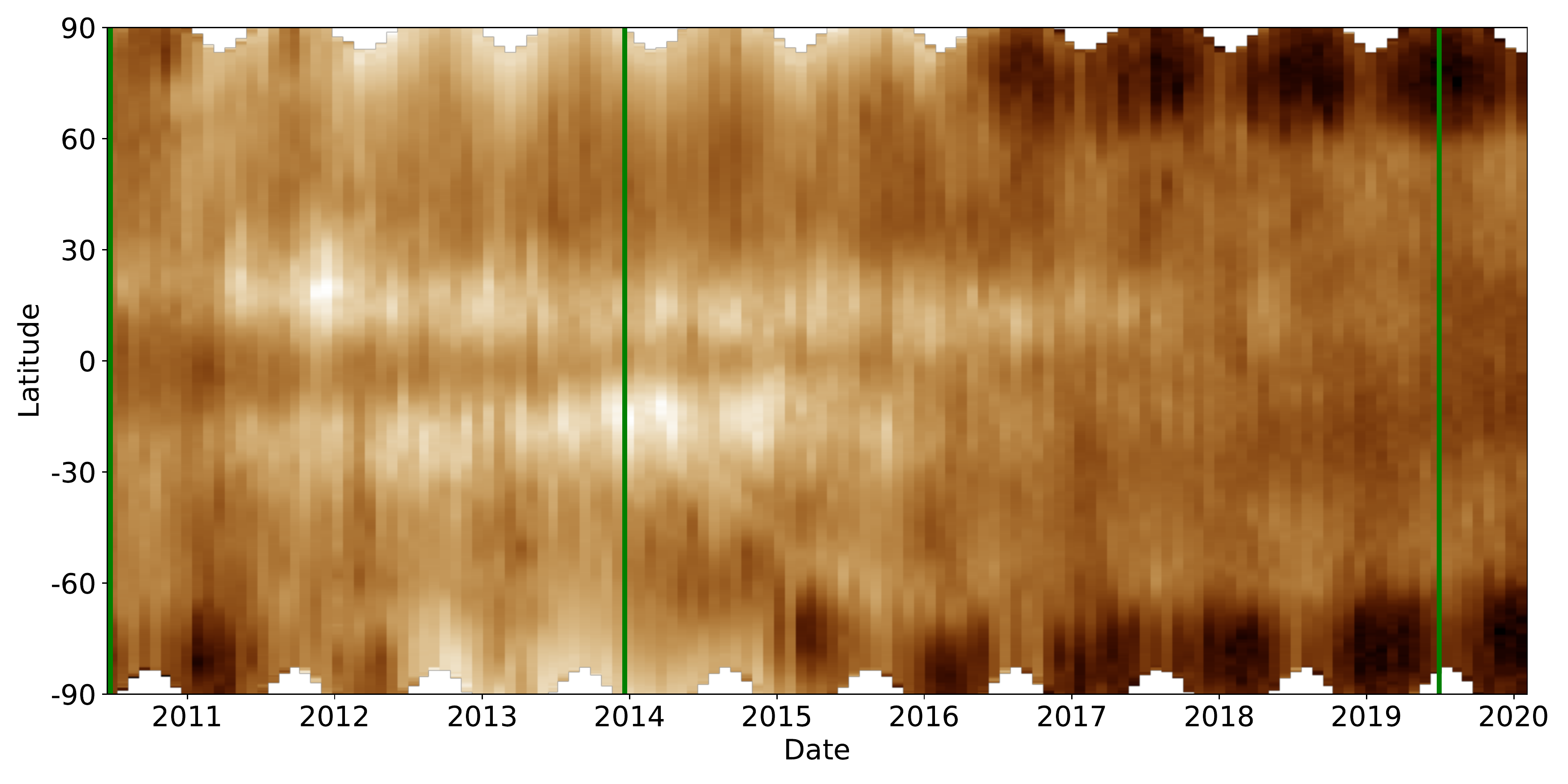}
\caption{Concatenation of synoptic maps averaged over
longitudes. Green vertical lines mark timestamps corresponding to CR~2098,
CR~2145, and CR~2219 shown in Figure~\ref{fig:maps}.
\label{fig:maps_stacked}}
\end{figure}

In Figure~\ref{fig:maps} we demonstrate examples of the constructed
synoptic maps for the solar activity maximum and minimum. Specifically,
for the demonstration in this and following figures, we choose three
Carrington Rotations: CR~2098 (during the solar minimum between Cycles 23 and 24),
CR~2145 (during the Cycle 24 maximum), and 
CR~2229 (during the minimum between Cycles 24 and 25).
One can notice that the synoptic
maps during solar minimum tend to be darker. This
is even clearer if we average the synoptic maps over 
longitude and concatenate them in the chronological order
(Figure~\ref{fig:maps_stacked}). Apart from the long-term
intensity variations associated with the solar cycle we also
find annual variations associated with the solar B0 angle,
best seen along a fixed latitude.
In our opinion, the latter effect can be related to the emitting plasma
that reduces the visibility of polar CHs observed close to the limb.
This effect has been discussed in \citet{Kirk_2009}.
The nature of cyclic variations is a matter of a separate
investigation. Another point that should be mentioned is
the instrument degradation issue. In our research, we employ
images provided by the SDO team, which are corrected for the
degradation effects, and do not apply additional processing.
However, the calibration process is not unique, and one could
consider alternative datasets, e.g. the one prepared by
\citet{Galvez_2019}.

To conclude this section we would like to mention that the source code
for synoptic maps construction is open-sourced in the GitHub repository
\url{https://github.com/observethesun/synoptic_maps}, while the synoptic maps produced for each Carrington rotation are available in a catalog \url{https://sun.njit.edu/coronal_holes/}.

\pagebreak
\section{Segmentation model} \label{sec:model}

We start with a brief description of the neural network model
proposed by \citet{Illarionov_2018} and discuss how to apply
it to the synoptic maps or, generally speaking, to input
images of arbitrary shape.

\begin{figure}[h!]
\centering
\includegraphics[width=1\textwidth]{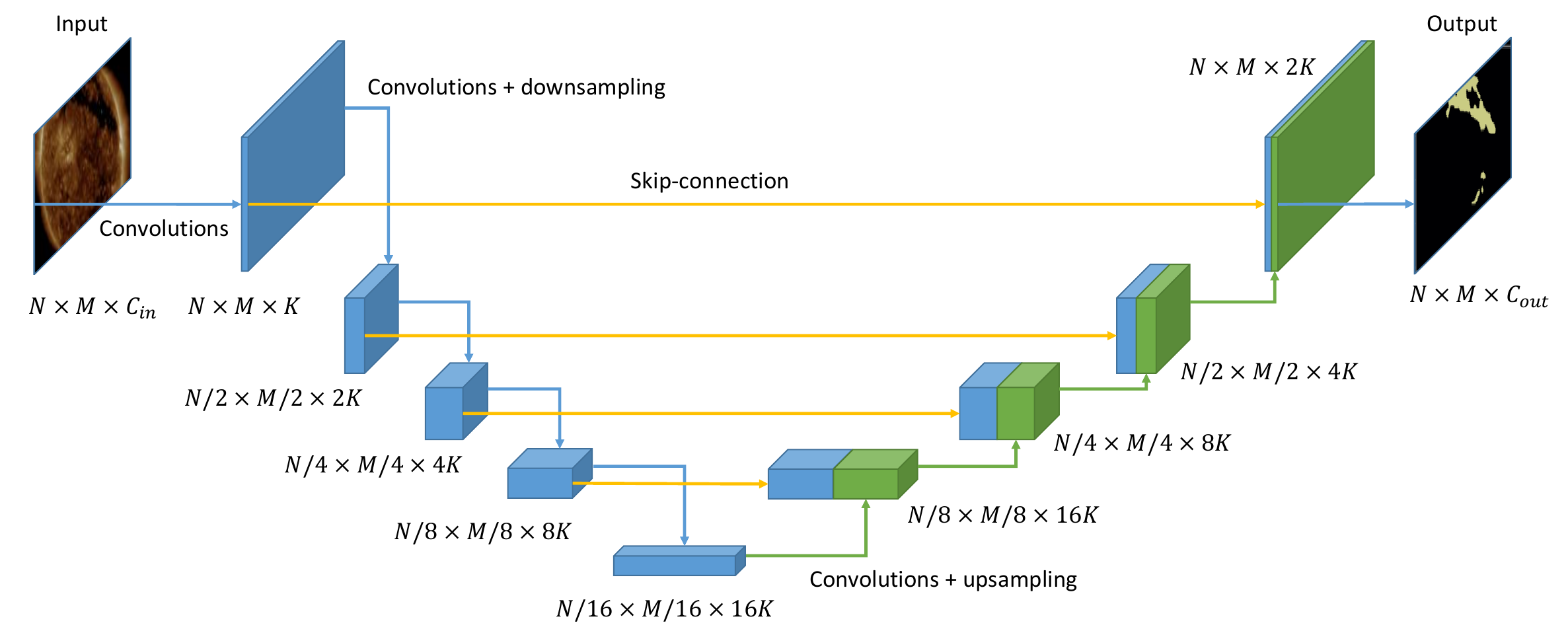}
\caption{The U-Net architecture with
compression and decompression branches and skip-connections.
The input images (e.g. solar disk image or synoptic map) have spatial
dimensions $N\times M$ and $C_{in}$ channels.
Each convolutional-downsampling block compresses spatial dimensions 
and increases the number of channels. The decompression branch acts as an inverse
operation, the output images (e.g. segmentation mask) have spatial dimensions
$N\times M$ and $C_{out}$ channels.
\label{fig:unet}}
\end{figure}

The model is a typical U-Net convolutional model \citep{unet}.
Figure~\ref{fig:unet} schematically shows the model architecture. 
It consists of two branches. The first branch compresses an input image
via a set of convolutional and downsampling operations into a tensor
with reduced spatial dimensions but an increased channel dimension.
Each downsampling operation reduces the spatial dimensions
by a factor of two, while each convolutional operation increases the number
of channels by the same factor of two. The number of the channels
after the first convolutional operation (denoted $K$ in Figure~\ref{fig:unet}) 
is a parameter of the model. The model we use has $K=24$.
It total, the compression branch consists of four convolutional-downsampling
steps. For example, for an input image of (256, 256) pixels and $K = 24$
the compression branch will result in a (16, 16, 384) tensor.

The second branch of the model is a decompression branch.
It consists of a set of convolutional-upsampling operations that,
simply speaking, act as an operation inverse to the compression branch.
The output image tensor will have the same dimensions as the input image. 
Because the compression-branch localization information
becomes more and more limited because of the downsamplings, the
U-Net architecture includes skip-connections between the
corresponding tensors in the compression and decompression
branches.  This operation stacks a copy of tensors in the
compression branch to tensors in the decompression branch.
Thus, layers in the decompression branch obtain information from
earlier layers with the localization information present.
Additional technical details of the implemented model can be found in the original 
paper \citep{Illarionov_2018}. The source code for the model application
to synoptic maps is available in the repository \url{https://github.com/observethesun/synoptic_maps}.

An important feature of the proposed model architecture is that it 
is independent of the input image shape. This means that the model 
can be trained on patches extracted from original images, and then
used for the analysis of full-size images. In this work, we 
apply the model trained on a set of disk images 
to the synoptic maps and pole-centric projections constructed
from these maps.

For the model training \citet{Illarionov_2018} 
used the binary masks of CHs obtained at the Kislovodsk Mountain
Astronomical Station. These binary masks separate CHs and non CHs regions, including flaments.
The binary masks along with other products are contained in
daily reports of the station. An archive of solar activity maps,
including CH boundaries is available at 
\url{https://observethesun.com}. Thus, the model training represents
a semi-automated and manually controlled process of the CH
identification applied at the station.

We use the same convolutional
kernels and other trainable parameters that were obtained
by \citet{Illarionov_2018}. This means that the presented results can be directly correlated
with the previous work.

There are some technical issues that we would like to mention.
First, the synoptic maps presented in Sec.~\ref{sec:maps} have the spatial 
resolution of $720\times360$ pixels. The model was trained on the $256\times256$-pixel 
disk images. Thus, it makes sense to downscale the synoptic maps
to better match the pixel sizes. Second, it is recommended to apply a maximal
intensity padding
to the synoptic maps to avoid some artifacts near the boundaries. The point is that due to the convolutional nature of the
model, each pixel of the next layer is connected only with a local
group of pixels in the previous layer and thus have a bounded
receptive field. It follows that the neurons in the deepest layer
classify pixels based on their local surroundings. Pixels near
image boundaries have fewer pixels around them in contrast to e.g.
pixels in the image center. In practice, we can see border
artifacts in segmentation output. Image padding is a common way to
overcome this problem. It can be shown that neurons at the end of
the compression branch have a receptive field of $140\times140$
pixels in the input image. Thus, additional padding of about 70
pixels around the synoptic map will provide a full receptive field
for pixels near synoptic map boundaries. Note that this action is
not required for the solar disk images since the space around the
solar disk acts as natural padding. We have tested various
approaches, e.g. constant, mean, and reflection padding, and find
that the most straightforward maximal intensity constant padding
works well.
To be more detailed, we downsample the synoptic maps to $360\times180$ pixels
and apply the spatial padding to obtain the target size of $512\times256$ pixels.
The CNN model applied to the $512\times256$ input images 
produces the segmentation masks of the same size from
which we extract a $360\times180$ region which 
contains the desired segmentation map for the synoptic map,
and is the final output.

\begin{figure}[h!]
\centering
\includegraphics[width=0.7\textwidth]{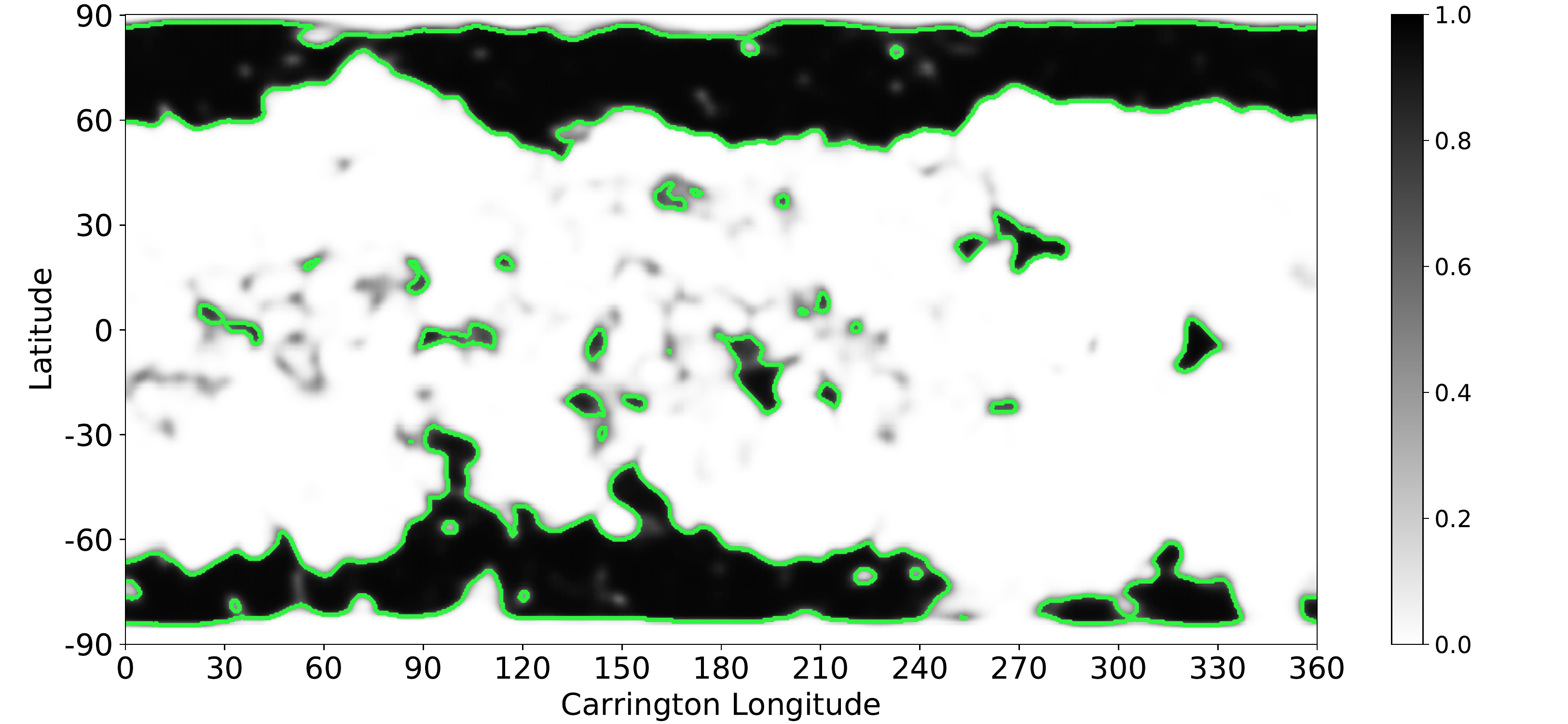}
\caption{A heatmap of CHs in the output of the CNN model for CR~2219 (shown in grayscale). Green lines correspond to a threshold
value of 0.5 used for binarization. In the following figures, we show only the boundaries of the binarized heatmaps.\label{fig:heatmap}}
\end{figure}

\begin{figure*}[h!]
\centering\includegraphics[width=0.8\textwidth]{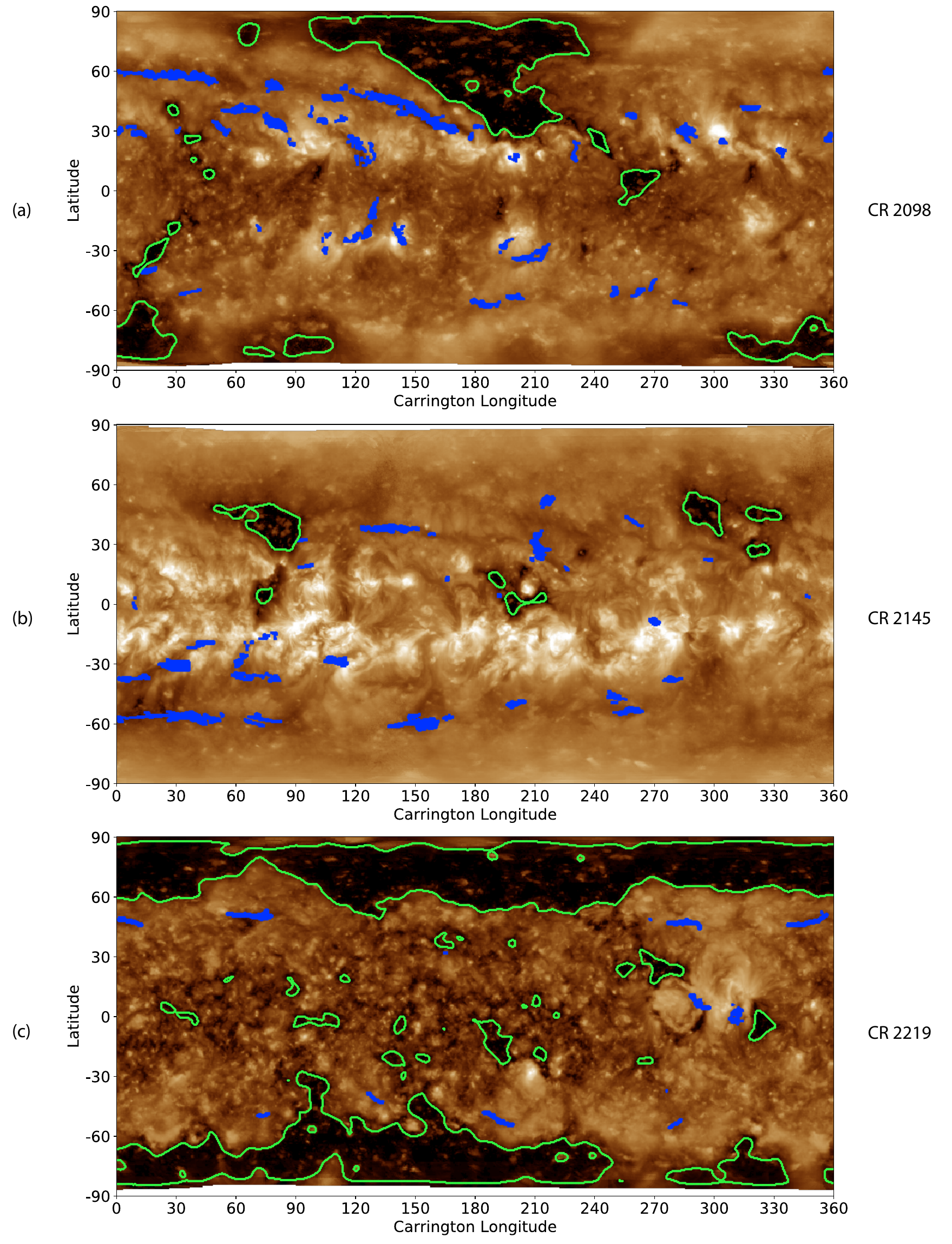}
\caption{Overlaid synoptic maps and reconstructed CH boundaries (green lines)
for CR~2098 (a), CR~2145 (b), and CR~2219 (c). These are the same CRs as in Figure~\ref{fig:maps}. 
For comparison, filaments from the catalogue of the Kislovodsk
Mountain Astronomical Station are shown in blue color.
\label{fig:pred}}
\end{figure*}

Figure~\ref{fig:heatmap} shows a sample segmentation map
obtained using the CNN model. The model outputs a
score for each pixel to be a part of a CH. The score ranges from 0 to 1.
We apply a 0.5 thresholding to convert the heatmaps into binary masks.
For example, Figure~\ref{fig:pred}
shows that the identified CHs boundaries correspond to visual expectation
and accurately detects CHs regions.
Moreover, we do not find misclassification 
examples with respect to the catalogue
of filaments provided by the Kislovodsk Mountain Astronomical Station and shown in blue color in the same plot.
In the next section, we provide a detailed analysis.

\begin{figure*}[h!]
\centering\includegraphics[width=\textwidth]{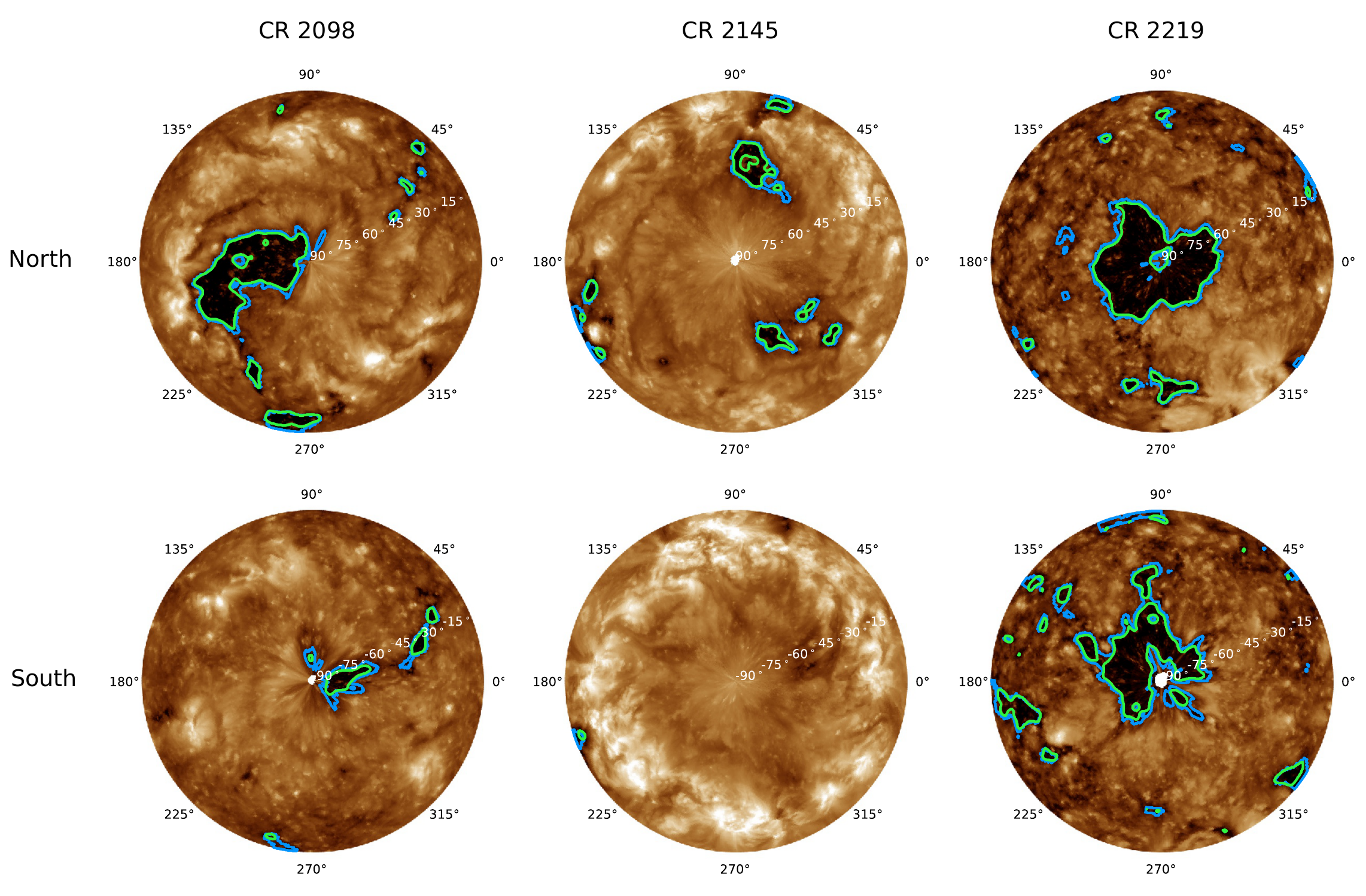}
\caption{CH boundaries (green lines) identified in the pole-centric input images (color background), in
comparison with the pole-centric projections of the CH boundaries deduced from the synoptic maps (blue lines).
Columns correspond to the same CRs as in Figure~\ref{fig:maps}. Top and bottom rows show the North and South pole projections.
\label{fig:circ}}
\end{figure*}

To demonstrate an additional application of the CNN model,
we apply it to the
pole-centric projections of the synoptic maps.
The model inference in this case is the same as for the solar disk
images. Figure~\ref{fig:circ} shows
sample segmentation maps obtained for the polar projection inputs.
For comparison, we put in the same figure pole-centric projections
of CHs obtained in synoptic maps. We note that both methods are
in  good agreement as it should be expected.

In Appendix, we discuss a possible
interpretation of the segmentation procedure
within the CNN model from a physical point of view.

\section{Analysis} \label{sec:results}

In this section, we demonstrate that the CH detection method is
stable against parameters of the construction of the synoptic maps,
and investigate general physical properties of CHs. 

The most essential parameter in the synoptic map construction is the strip
width (in our notation it is represented by the shift and scale parameters).
Indeed, the wider strips result in smoother maps without finer details,
while narrower strips preserve details but provide noisier maps.
Another point is that due to the limb-brightening effect the strip
width also affects the pixel intensity distribution. To avoid
this effect we apply the histogram matching procedure as described in
Sec.~\ref{sec:maps}.

\begin{figure}[h!]
\centering\includegraphics[width=1\textwidth]{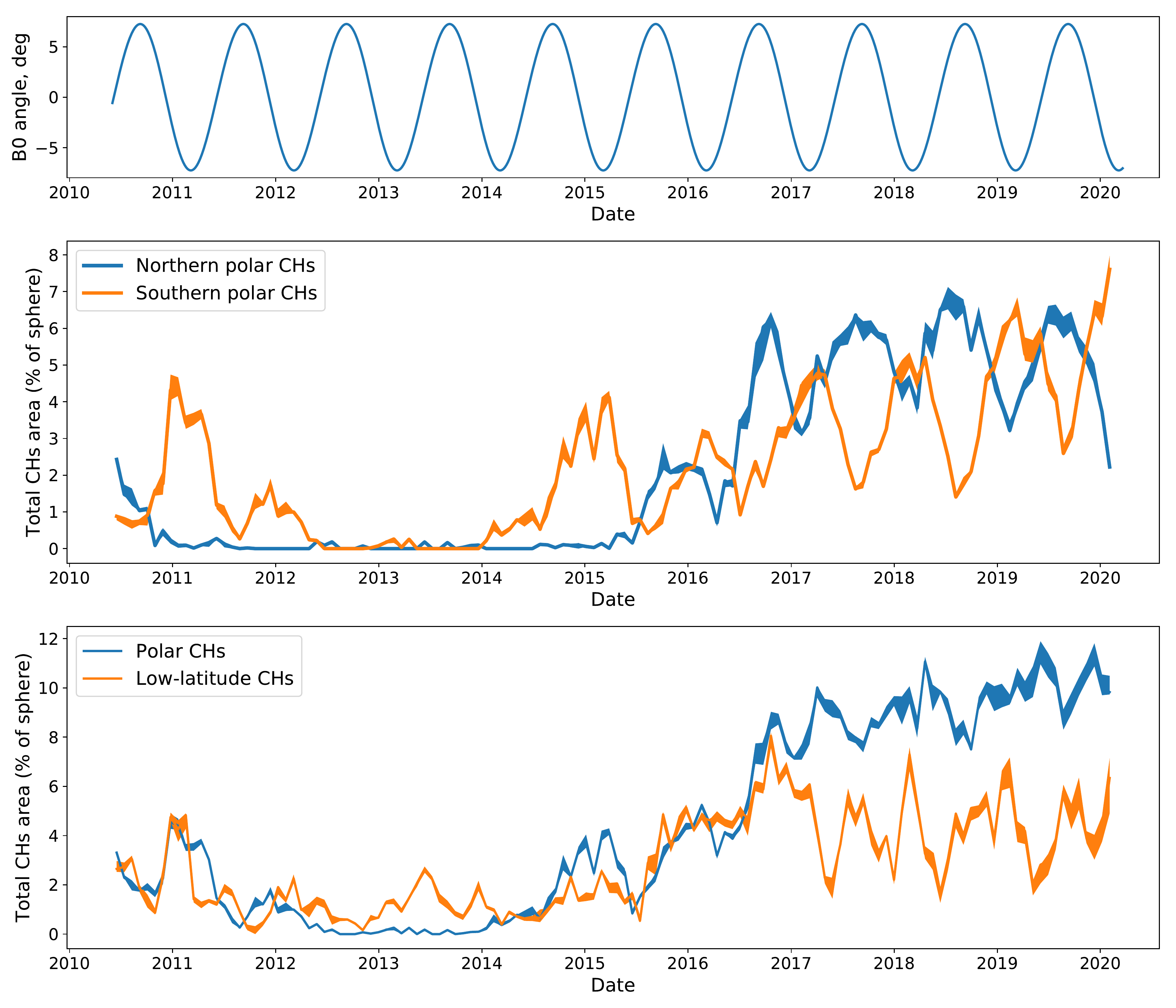}
\caption{Upper panel: yearly variation of the solar B0 angle. Middle panel: areas of the northern and southern polar CHs.
Bottom panel: areas of the polar and low-latitude CHs. The separating
boundary between polar and low-latitude regions
is $\theta = \pm 50^{\circ}$. Line width corresponds to
uncertainties that arise from different parameters of synoptic
maps construction.
\label{fig:area_ns}}
\end{figure}

For the uncertainty estimation we consider all combinations of
values of the shift parameter: \{$6.6^{\circ}$, $13.2^{\circ}$,
$19.8^{\circ}$, $26.4^{\circ}$, $33.0^{\circ}$, $39.6^{\circ}$\}
and the scale parameter: \{0.5, 1, 2, 4\}.
Note that the extreme cases correspond approximately to the
narrowest possible strip (about $\pm 6.6^{\circ}$ around
the central meridian with a thin blending zone), and a case
where each pixel of the synoptic map results from averaging of
6 nearest disk images. In Figure~\ref{fig:area_ns} we show
intervals between the smallest and largest total areas
obtained for all parameter combinations.
One can notice that the uncertainties are rather negligible. 
This important point allows us to conclude that the CH regions 
detected in the synoptic maps do not depend on a particular 
way of the map compilation, but represent stable and physical structures.

Figure~\ref{fig:area_ns} shows the CH areas as a function of time
separately for the Northern and Southern hemispheres as well as for 
the polar ($|\theta| > 50^{\circ}$) and low-latitude
($|\theta| \le 50^{\circ}$) zones.
Our choice of separating boundary $\theta = \pm 50^{\circ}$ 
is consistent with the work of \citet{Webber_2014}. 
We take into account the contribution
of individual pixels into each of these groups rather
than attribute a whole CH based on the location of its center. 
Thus, pixels from the same CH may
contribute to the different groups.
We make several observations from the figure.
First, large annual variations seen in the middle panel have a
clear connection to the variations of the solar B0 angle shown in
the upper panel (due to its variations, the North and South poles
of the Sun are alternately hidden from the observations). Peaks of
both lines in the middle panel correspond to the maximal absolute
values of B0 when the North or South poles are best seen. Second,
there is an asymmetry
between the North and South. We observe the hemispheric asymmetry both in time
(the area of the Southern polar CHs decreases later and starts to
increase earlier than the area of the Northern CHs) and in amplitude
(the southern polar CHs demonstrate increasing trend during the
solar minimum between Cycles 24 and 25, while the northern CHs do
not show this trend). 
\citet{Webber_2014} also demonstrated asymmetries
in the polar CHs during the solar minimum between Cycles 23 and 24.
Third, from the bottom panel, we find that the solar minimum manifests
itself in increasing both the polar and low-latitude areas of CHs.
Moreover, while the areas of the polar CHs continue to increase,
the low-latitude CH areas fluctuate near constant value. 
This may be consistent with ideas of the solar flux transport theory 
that magnetic fields migrate from low-latitudes to
the poles and accumulate there during solar minimums
\citep[see][]{Babcock1961, Leighton1969}.

\begin{figure}[h!]
\centering\includegraphics[width=1\textwidth]{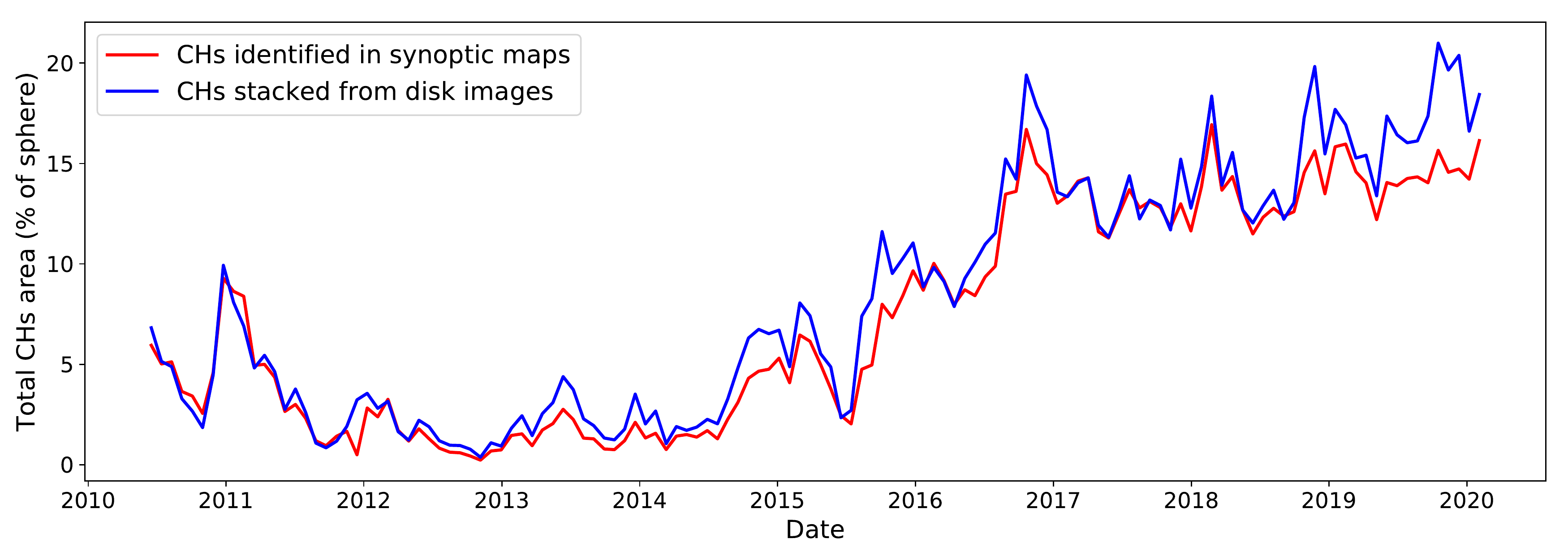}
\caption{Total area of CHs identified in synoptic maps (red line) in comparison to CHs stacked into synoptic maps from CHs identified in solar disk images (blue line).
\label{fig:area_disk}}
\end{figure}

As we noted before, synoptic maps are not directly
observable data in contrast to solar disk images. Thus it is
interesting to compare CHs identified in disk images and synoptic
maps. To make this comparison feasible we apply the CNN model to
solar disk images as described in \citet{Illarionov_2018} and
stack obtained binary masks of CHs into synoptic maps.
To construct the synoptic maps from binary
CHs masks we apply the same procedure as 
for solar disk images excluding the histogram
matching step. Figure~\ref{fig:area_disk}
shows total area of CHs identified in solar disk images and
stacked into synoptic map in comparison to CHs identified
directly in synoptic maps. In should be noted that production
of synoptic maps from binary masks is much more sensible to
the shift and scale parameters in comparison to production
of synoptic maps from disk images. The point is that the
narrower the strips are, the more noisy map we obtain.
We set the shift parameter to $39.6^{\circ}$ and scale to 4
to ensure that only stable structures identified in disk
images contribute to synoptic maps. We find that this
choice of parameters provides the best correlation
with CHs identified directly in synoptic maps.
Thus we conclude from Figure~\ref{fig:area_disk}
that CHs identification in solar disk images and synoptic
maps is in agreement.

\begin{figure}[h!]
\centering\includegraphics[width=0.7\textwidth]{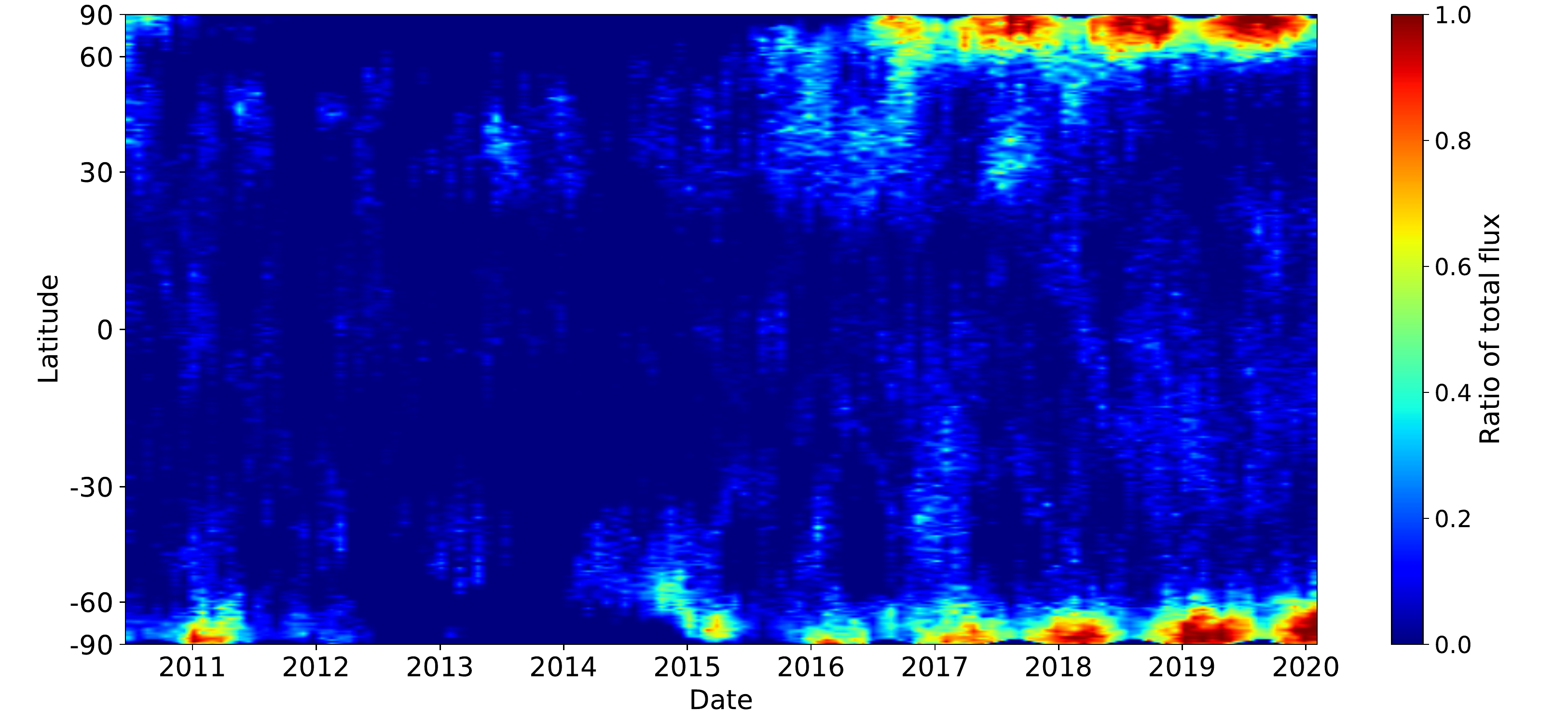}
\caption{Ratio of total unsigned magnetic flux in CHs to the total unsigned magnetic flux
integrated over all longitudes.
\label{fig:flux_ratio}}
\end{figure}

Now we consider synoptic maps of CHs with respect to 
magnetic synoptic maps and construct time-latitude and
time-longitude diagrams.
We start with the time-latitude diagram that shows a ratio of total
unsigned magnetic flux in CHs to the total unsigned magnetic flux
integrated over all longitudes (Figure~\ref{fig:flux_ratio}).
We conclude from this plot that while the solar minimum is 
accompanied by an increase of the low-latitudes CHs areas (see
Figure~\ref{fig:area_ns}, lower panel), its contribution
to the total unsigned flux is not dominant. In contrast,
polar CHs generate almost the whole unsigned magnetic flux.
Note that for construction of this
plot we thresholded  unsigned magnetic synoptic maps at 10 Gauss to
avoid noise contribution.

\begin{figure}[h!]
\plotone{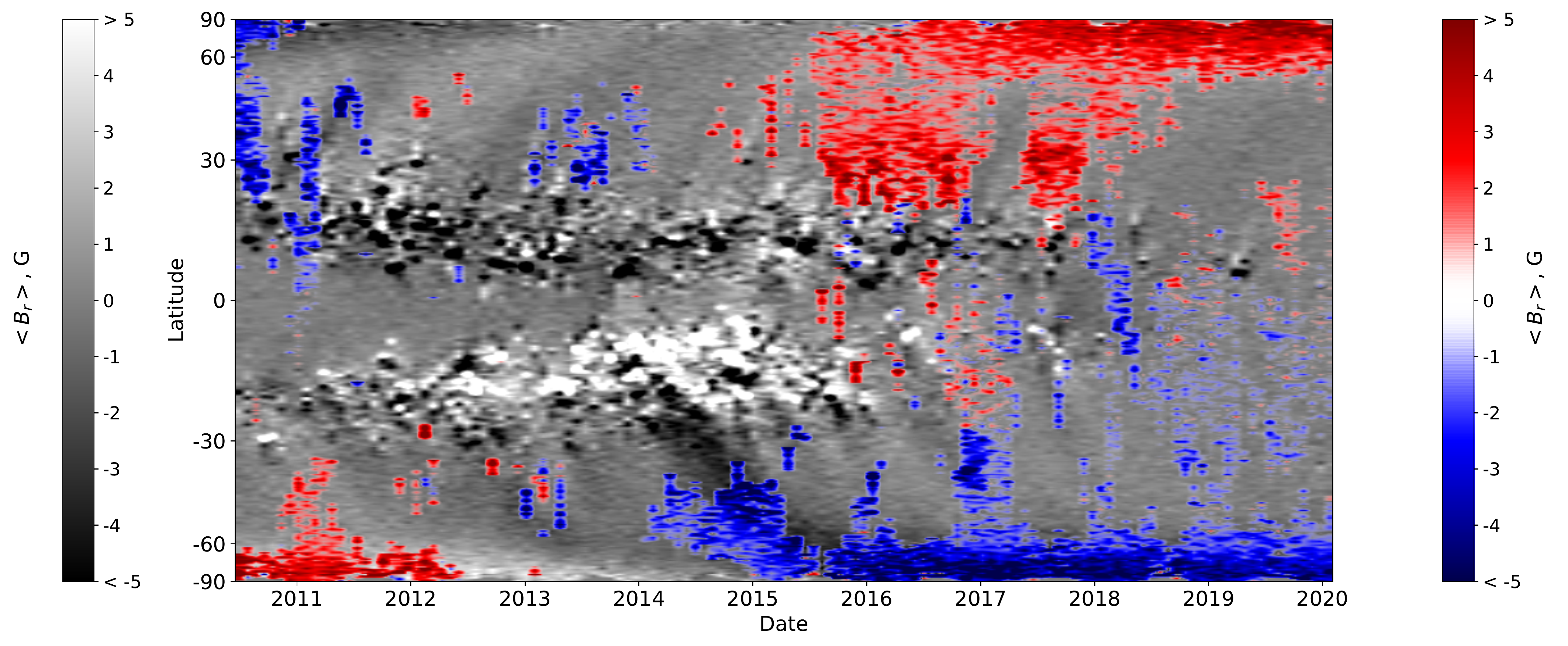}
\caption{Time-latitude diagram of the longitudinally averaged magnetic field in CH regions
(shown in blue and reds colors) and magnetic field averaged over all longitudes (grayscale map).
Note that neutral color in red-blue color bar is not white but
transparent so that weak CH magnetic fields are not
visible in the plot.
\label{fig:flux_Br}}
\end{figure}

For a more detailed investigation, we take into account the sign
of the magnetic field. In Figure~\ref{fig:flux_Br}, the grayscale 
background is a magnetic field averaged over all longitudes
while blue and red colors show the magnetic field averaged over longitudes
only in CH regions.
Note that averaging CHs magnetic
field we filter out latitudes
where CHs cover less than $20^{\circ}$
of longitudes in total to prevent plotting of statistically insignificant values. We find from this plot that
polar latitudes have a prevalent sign
of the magnetic field that is opposite
in North and South and between solar cycles.
Also in agreement with Figure~\ref{fig:flux_ratio},
we find that CHs at lower latitudes in the minimum
between Cycles 24 and 25 have significantly
lower magnetic fields in contrast to polar CHs.

A detailed investigation of results presented in Figure~\ref{fig:flux_Br}
can give insights about the origin of the CHs open magnetic flux and its relation to the flux-transport mechanism.
For example, \citet{Golubeva_2017} associated CHs
with decaying complexes of magnetic activity, while studies of \citet{Tlatov_2014} and \citet{Huang_2017} revealed pole-to-pole open flux migration. \citet{Hamada_2018} presented a similar plot showing dominant polarity and relative areas of CHs for Cycles 23 and 24. To facilitate we have constructed the CH catalog and made it publicly available. 

\begin{figure*}[h!]
\centering\includegraphics[width=1\textwidth]{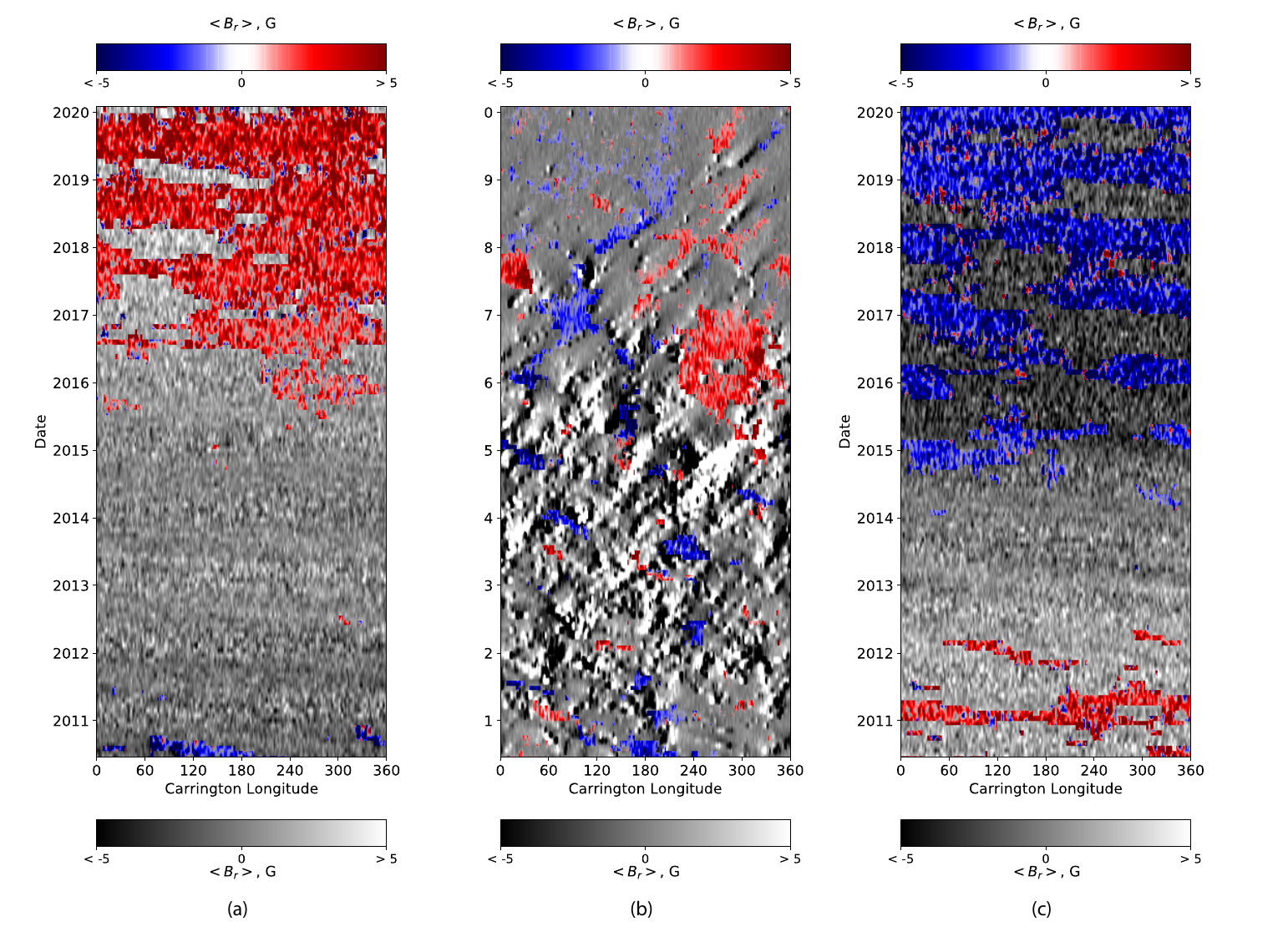}
\caption{Time-longitude diagrams of CH magnetic fields 
(shown in blue and red colors) in three latitudinal
zones. Panel (a) is for high latitudes in the Northern hemisphere
($\theta > 50^{\circ}$), panel (b) is for lower latitudes
($|\theta| \le 50^{\circ}$), panel (c) is for high latitudes in the
Southern hemisphere ($\theta < -50^{\circ}$). The grayscale 
background shows the magnetic field averaged over latitudes for each latitudinal zone.
Note that neutral color in red-blue color bar is not white but
transparent so that weak CH magnetic fields are not
visible in the plot.
\label{fig:flux_long}}
\end{figure*}

Finally, we demonstrate time-longitude diagrams of the CH magnetic fields.
Panels in Figure~\ref{fig:flux_long} correspond to three regions located at 
northern polar latitudes, low-latitudes, and southern polar latitudes.
The separating boundaries are $\theta = \pm 50^{\circ}$ as in
Figure~\ref{fig:area_ns}. We observe that CHs patterns
are substantially different in the high and low latitude regions.
At the high latitudes, we find large-scale structures that
exist for about a year. This indicates that
CHs form stable sector structures in the magnetic field distribution.
In the low latitude region, we find a mixture of two populations.
Before 2015 (during the solar maximum) one can observe small-scale structures
that exist for several months. 
After 2015 (during the solar minimum)
we find characteristics strip structures that can be traced for several years.
A final remark from Figure~\ref{fig:flux_long} is about
the inclination of the structures across all there panels.
The elongation from the bottom right to the top left (which we see
in the high-latitude zones) means that
the region rotates slower than the Carrington coordinate
system. In contrast, the opposite elongation at the low latitudes
means the faster rotation. This is consistent with 
the general picture of the  differential rotation of the Sun. 
However, a detailed analysis and rotation speed estimation 
is out of the scope of this paper.

\section{Conclusions}

We have demonstrated that a Convolutional Neural Network (CNN) 
model trained to identify CHs in the solar disk images is capable
to detect CHs in the solar synoptic maps without any
additional adjustments. Being composed of only convolutional
operations the CNN processes images of any shape
in the same way. This also implies that the local image
content dominates over the global content (i.e. the segmentation
result will be the same for portions of the image and
the whole image). Due to these facts, one can expect
that for CNN it should be the same whether it sees
the whole disk image, a partial disk image, or a synoptic map
(we suppose that human interpretation acts similarly).

To illustrate this idea, we constructed a dataset of
synoptic maps from daily solar disk images
used for model training. The process of synoptic map construction
is not unique and contains free parameters. We have
shown that the segmentation procedure is stable for
a wide range of parameter values (Figure~\ref{fig:area_ns}).

It is not trivial to compare properties of CHs identified 
in the disk images and synoptic maps, because it requires a 
construction of the binary synoptic maps from binary 
segmentation masks of the  disk images. However, there is
a more feasible option. One can build pole-centric
projections of the synoptic maps, make a segmentation
using the CNN model, and compare the output with the pole-centric
projections of the binary synoptic maps. For a proper segmentation
model, the results should be in agreement. Indeed, in 
Figure~\ref{fig:circ} we find that the CH boundaries
obtained by both methods are very close. Thus, we 
conclude that the CNN model recognizes CHs regardless
of the way we project them. In other words, it learns
what a CH is itself rather than how a CH looks in
the solar disk context.

For our initial investigation of the physical properties of the CHs, 
we separated them into polar and low-latitudinal,
and also into northern and southern.
In Figure~\ref{fig:area_ns} we find that 
the CH areas are minimal during the solar maximum
and start to increase during the declining phase of the solar cycle.
There are visible asymmetries between the North and South
both in the temporal behavior and in the magnitude of CH areas.

Finally,  in Figures~\ref{fig:flux_ratio}, \ref{fig:flux_Br},
and \ref{fig:flux_long} we demonstrated magnetic field patterns associated with CHs in the time-latitude and time-longitude domains. In Figure~\ref{fig:flux_Br}, we compare the CH patterns with the longitudinally averaged magnetic synoptic maps (so-called magnetic “butterfly” diagram). The magnetic butterfly diagram reveals the transport of magnetic flux of decaying active regions from the low and mid latitudes to the polar regions. As we mentioned above, it was previously suggested that the CHs are formed at high latitudes from the magnetic field associated with the flux transport events. Figure~\ref{fig:flux_Br} shows that this association is not common. In some cases, e.g. in the Southern hemisphere around 2015, 2016 and 2017 we can see the association of CHs of negative polarity with the negative flux transport. In particular, in the Southern hemisphere the most prominent zone of CH formation, which was around 2015, partially overlaps with a major flux transport event (Fig.~\ref{fig:flux_Br}). In the Northern hemisphere, the CH activity was a year later and lasted longer, in 2016--17. It is not so apparently associated with the flux-transport events. This zone was also compact in the Carrington longitude, located around 240--300 degrees (Fig.~\ref{fig:flux_long}b). A major complex of activity was in the zone, but a year earlier. On the other hand, CHs of the southern (negative) polarity were more scattered in longitude. Perhaps, as shown in a case study by \citet{Benevolenskaya2012}, CHs can also be associated with magnetic fields emerging at high latitudes from the sub-photospheric layers. The relationship between the CHs and magnetic flux emergence and transport requires further detailed investigation.

Thus, our research demonstrates that CNN is a powerful
and flexible tool for the investigation of solar activity.
In particular, it enables a unified approach to the
identification and characterization of CHs in various 
geometrical representations of solar image data. 
To make this approach more readily available, we open-sourced
the code for synoptic map construction and CHs segmentation
in the repository \url{https://github.com/observethesun/synoptic_maps} and
opened free access to CHs synoptic maps in the catalog \url{https://sun.njit.edu/coronal_holes/}
available in FITS and JPEG formats.

\acknowledgments
We thank the reviewer for valuable comments and suggestions.

The work was partially supported by RSF grant 20-72-00106, RFBR grant 18-02-00098-a, NSF grants 1639683, 1743321, 1927578, and
NASA grants  80NSSC19K0630, 80NSSC20K0602.

\appendix

Here we provide some insights about how the proposed CNN model works.
We stress that this discussion is only an interpretation rather than an explanation.
Nevertheless, it helps to reveal a physical basis for the produced segmentation
maps.

\begin{figure*}[h!]
\centering\includegraphics[width=1\textwidth]{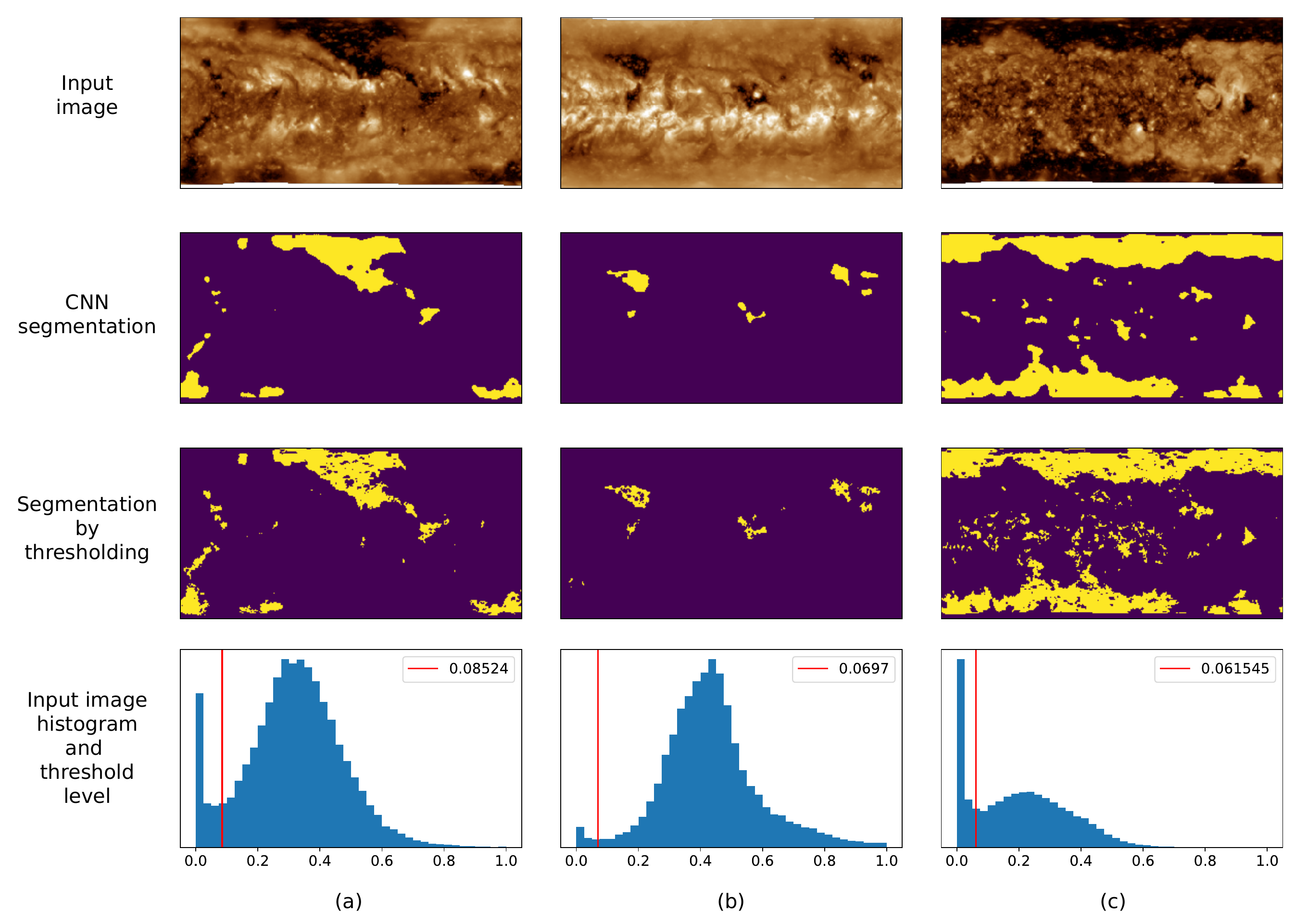}
\caption{Comparison of CNN and threshold-based segmentation.
Columns correspond to CR~2098 (a), CR~2145 (b), and CR~2219 (c). 
These are the same CRs as in  Figure~\ref{fig:maps}.
Top row shows the input synoptic maps. Second row demonstrates the CHs segmentation
by the CNN model. Third row shows the equivalent, in terms of CHs pixel counts,
threshold-based segmentation. Bottom row shows histograms of
the pixel intensity distributions and the threshold levels which 
provide the equivalent segmentation in terms of the pixel counts. 
\label{fig:thresholds}}
\end{figure*}

A typical alternative to the CNN segmentation is a threshold-based segmentation.
The most straightforward approach is to select some threshold level
for pixel intensities and declare everything beyond this level as CHs.
It is interesting to investigate to what extent the CNN model is more advanced.
In our experiments, we consider several synoptic maps
(same as in Figure~\ref{fig:maps}) and determine the threshold levels that
result in the same number of pixels corresponding to CHs as in
the segmentation masks from the CNN model. We stress that we only match
pixel counts while finding the thresholds. In Figure~\ref{fig:thresholds}
we demonstrate the input synoptic maps, the CH segmentation maps produced by
the CNN model, and equivalent (in the sense of the CH pixel counts) segmentation maps 
produced by the thresholding. The histograms show pixel intensity
distributions in the synoptic maps and the threshold levels. We make 
several observations from these plots. The CNN segmentation maps look
less noisy compared to the threshold-based segmentation. This means 
that the CNN acts not as a thresholding procedure but includes some high-level
processing. The second and more important observation is that the equivalent 
threshold in the CNN segmentation varies from image to image. 
This means that it depends on the image context.
 However, from a physical point of view, the most interesting note 
 is that the equivalent threshold
corresponds to the first minimum in the intensity distribution.
Most of the CHs segmentation algorithms proposed
earlier rely on this idea more or less explicitly. In this
respect, the CNN model automatically finds this more or less 
reasonable and intuitive strategy.

\begin{figure*}[h!]
\centering\includegraphics[width=1\textwidth]{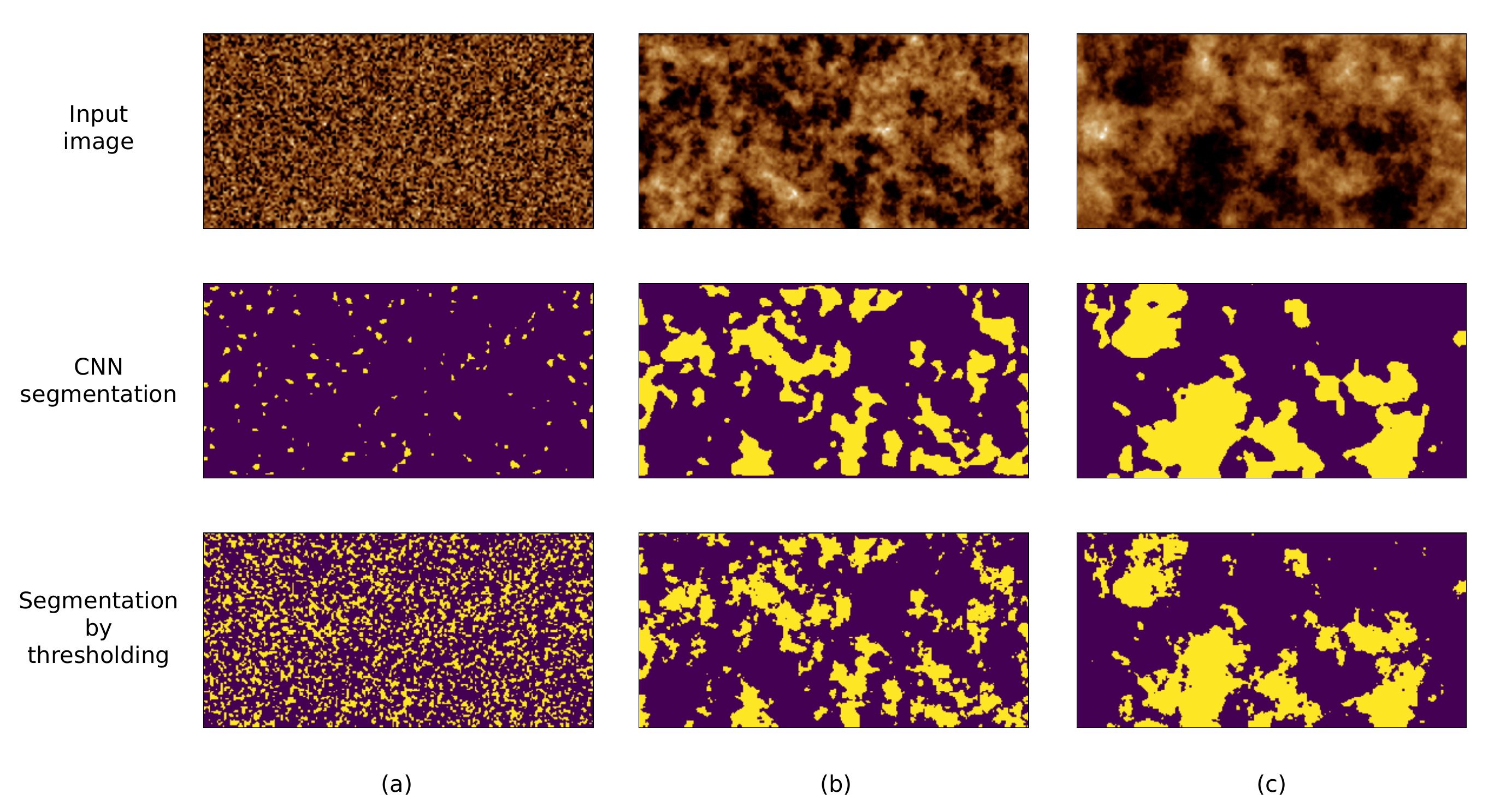}
\caption{Sample of synthetic synoptic maps and
corresponding segmentation maps.
Columns correspond to the correlation radius parameter
$r_0$ = 0.01 (a), 10 (b) and 20 (c). Top row
shows the synthetic synoptic maps. Middle row shows the segmentation maps
obtained using the CNN model. Bottom row shows the threshold-based 
segmentation.
\label{fig:samples}}
\end{figure*}

Now we want to take a step deeper and consider some
synthetic cases. We noted in Figure~\ref{fig:thresholds} that
while being equivalent in terms of the CH pixel counts to the thresholding
procedure, the CNN segmentation masks are not as noisy as the 
threshold-based ones. To investigate this fact in more detail, 
we generate a set of synthetic synoptic maps as Gaussian random structures 
with radial exponential correlation function $K(r) = \exp(-r/r_0)$. Here $r$ is
a distance between pixels in the pixel units, and $r_0$ is a
correlation radius.
Varying $r_0$ we obtain a set of synthetic maps ranging from
the maps with almost uncorrelated noise for small $r_0$
to the maps with large-scale correlated random structures 
for large $r_0$. For each map,
we apply the histogram matching procedure and make its
distribution similar to the solar synoptic map corresponding to CR 2219
(see the right column in Figure~\ref{fig:thresholds}).
Thus, for the threshold-based approach,
each synthetic map contains the same number
of pixels assigned to CHs (the threshold is also the same
as for the synoptic map corresponding to CR 2219).
Our goal is to compare this against the CNN model.
In fact, we vary $r_0$ from 0.01 to 20 and use a sample
of ten synthetic maps for each $r_0$.
Figure~\ref{fig:samples} shows a sample of the synthetic map for 
various $r_0$ and the corresponding segmentation maps.

\begin{figure}[h!]
\centering
\includegraphics[width=0.6\textwidth]{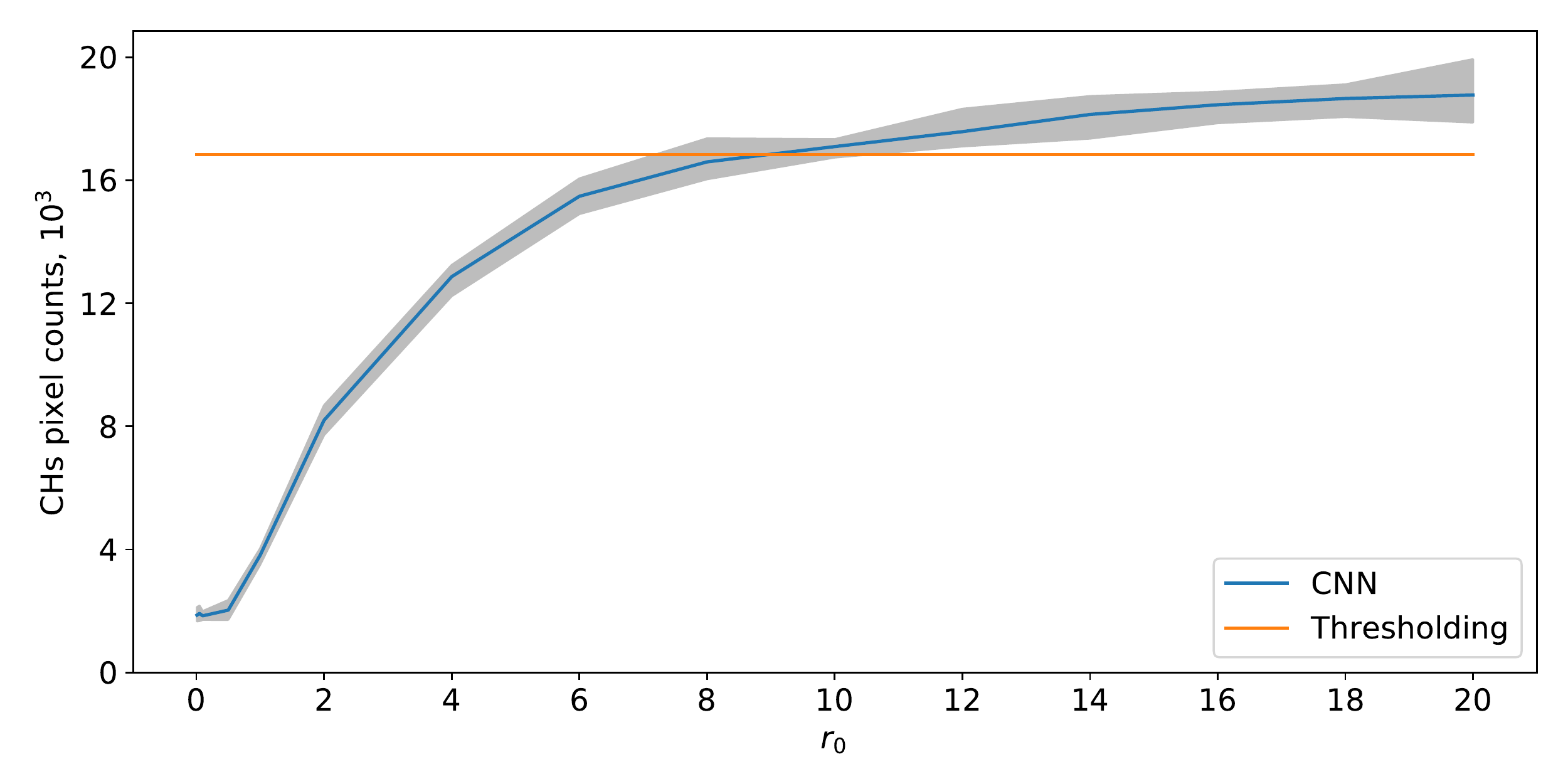}
\caption{CH pixel counts for the CNN model (blue line) 
and the thresholding method (orange line). Horizontal axis shows
the correlation radius $r_0$ used for synthetic synoptic map
sampling. Gray color shows a min-max range within 10 samples.
\label{fig:counts}}
\end{figure}

We note in Figure~\ref{fig:samples} that both segmentation
methods give mostly similar results for large-scale structures, but
substantially differ for small-scale structures. This is
also a reasonable feature of the CNN model trained for the CH
segmentation. Indeed, CHs are typically large-scale structures
so a proper model should take into account the size factor.
While for a typical CH segmentation method a region filtering 
procedure is an explicit part of the algorithm,
for the CNN model this step works automatically.
Figure~\ref{fig:counts} demonstrates the number of pixels
labeled as CHs against the scale factor (or the correlation radius
$r_0$ in our notations). Note that for the threshold-based
segmentation the pixel count is a constant because each
synthetic synoptic map has the same intensity distribution.

\bibliography{references}{}

\begin{thebibliography}{}
\expandafter\ifx\csname natexlab\endcsname\relax\def\natexlab#1{#1}\fi
\providecommand{\url}[1]{\href{#1}{#1}}
\providecommand{\dodoi}[1]{doi:~\href{http://doi.org/#1}{\nolinkurl{#1}}}
\providecommand{\doeprint}[1]{\href{http://ascl.net/#1}{\nolinkurl{http://ascl.net/#1}}}
\providecommand{\doarXiv}[1]{\href{https://arxiv.org/abs/#1}{\nolinkurl{https://arxiv.org/abs/#1}}}

\bibitem[{Abramenko {et~al.}(2009)Abramenko, Yurchyshyn, \&
  Watanabe}]{Abramenko2009}
Abramenko, V., Yurchyshyn, V., \& Watanabe, H. 2009, Solar Physics, 260, 43,
  \dodoi{10.1007/s11207-009-9433-7}

\bibitem[{{Babcock}(1961)}]{Babcock1961}
{Babcock}, H.~W. 1961, \apj, 133, 572, \dodoi{10.1086/147060}

\bibitem[{{Benevolenskaya}(2012)}]{Benevolenskaya2012}
{Benevolenskaya}, E.~E. 2012, Geomagnetism and Aeronomy, 52, 829,
  \dodoi{10.1134/S0016793212070031}

\bibitem[{{Caplan} {et~al.}(2016){Caplan}, {Downs}, \& {Linker}}]{Caplan_2016}
{Caplan}, R.~M., {Downs}, C., \& {Linker}, J.~A. 2016, \apj, 823, 53,
  \dodoi{10.3847/0004-637X/823/1/53}

\bibitem[{Cranmer(2009)}]{Cranmer_2009}
Cranmer, S.~R. 2009, Living Reviews in Solar Physics, 6,
  \dodoi{10.12942/lrsp-2009-3}

\bibitem[{Galvez {et~al.}(2019)Galvez, Fouhey, Jin, Szenicer,
  Mu{\~{n}}oz-Jaramillo, Cheung, Wright, Bobra, Liu, Mason, \&
  Thomas}]{Galvez_2019}
Galvez, R., Fouhey, D.~F., Jin, M., {et~al.} 2019, The Astrophysical Journal
  Supplement Series, 242, 7, \dodoi{10.3847/1538-4365/ab1005}

\bibitem[{{Garton} {et~al.}(2018){Garton}, {Gallagher}, \&
  {Murray}}]{Garton_2018}
{Garton}, T.~M., {Gallagher}, P.~T., \& {Murray}, S.~A. 2018, Journal of Space
  Weather and Space Climate, 8, A02, \dodoi{10.1051/swsc/2017039}

\bibitem[{Golubeva \& Mordvinov(2017)}]{Golubeva_2017}
Golubeva, E.~M., \& Mordvinov, A.~V. 2017, Solar Physics, 292,
  \dodoi{10.1007/s11207-017-1200-6}

\bibitem[{Gonzalez \& Woods(2006)}]{Gonzalez_2006}
Gonzalez, R.~C., \& Woods, R.~E. 2006, Digital Image Processing (3rd Edition)
  (USA: Prentice-Hall, Inc.)

\bibitem[{Hamada {et~al.}(2020)Hamada, Asikainen, \& Mursula}]{Hamada_2020}
Hamada, A., Asikainen, T., \& Mursula, K. 2020, Solar Physics, 295,
  \dodoi{10.1007/s11207-019-1563-y}

\bibitem[{Hamada {et~al.}(2018)Hamada, Asikainen, Virtanen, \&
  Mursula}]{Hamada_2018}
Hamada, A., Asikainen, T., Virtanen, I., \& Mursula, K. 2018, Solar Physics,
  293, \dodoi{10.1007/s11207-018-1289-2}

\bibitem[{Harvey \& Recely(2002)}]{Harvey_2002}
Harvey, K., \& Recely, F. 2002, Solar Physics, 211, 31,
  \dodoi{10.1023/A:1022469023581}

\bibitem[{Heinemann {et~al.}(2019)Heinemann, Temmer, Heinemann, Dissauer,
  Samara, JerÄiÄ‡, Hofmeister, \& Veronig}]{Heinemann_2019}
Heinemann, S., Temmer, M., Heinemann, N., {et~al.} 2019, Solar Physics, 294,
  144, \dodoi{10.1007/s11207-019-1539-y}

\bibitem[{{Henney} \& {Harvey}(2005)}]{Henney_2005}
{Henney}, C.~J., \& {Harvey}, J.~W. 2005, in Large-scale Structures and their
  Role in Solar Activity, Vol. 346, 261.
\newblock \doarXiv{astro-ph/0701122}

\bibitem[{{Hess Webber} {et~al.}(2014){Hess Webber}, {Karna}, {Pesnell}, \&
  {Kirk}}]{Webber_2014}
{Hess Webber}, S.~A., {Karna}, N., {Pesnell}, W.~D., \& {Kirk}, M.~S. 2014,
  \solphys, 289, 4047, \dodoi{10.1007/s11207-014-0564-0}

\bibitem[{{Huang} {et~al.}(2017){Huang}, {Lin}, \& {Lee}}]{Huang_2017}
{Huang}, G.~H., {Lin}, C.~H., \& {Lee}, L.~C. 2017, Scientific Reports, 7,
  9488, \dodoi{10.1038/s41598-017-09862-2}

\bibitem[{Illarionov \& Tlatov(2018)}]{Illarionov_2018}
Illarionov, E.~A., \& Tlatov, A.~G. 2018, Monthly Notices of the Royal
  Astronomical Society, 481, 5014, \dodoi{10.1093/mnras/sty2628}

\bibitem[{{Karna} {et~al.}(2014){Karna}, {Hess Webber}, \&
  {Pesnell}}]{Karna_2014}
{Karna}, N., {Hess Webber}, S.~A., \& {Pesnell}, W.~D. 2014, \solphys, 289,
  3381, \dodoi{10.1007/s11207-014-0541-7}

\bibitem[{{Kirk} {et~al.}(2009){Kirk}, {Pesnell}, {Young}, \& {Hess
  Webber}}]{Kirk_2009}
{Kirk}, M.~S., {Pesnell}, W.~D., {Young}, C.~A., \& {Hess Webber}, S.~A. 2009,
  \solphys, 257, 99, \dodoi{10.1007/s11207-009-9369-y}

\bibitem[{{Krista} \& {Gallagher}(2009)}]{Krista2009}
{Krista}, L.~D., \& {Gallagher}, P.~T. 2009, \solphys, 256, 87,
  \dodoi{10.1007/s11207-009-9357-2}

\bibitem[{{Leighton}(1969)}]{Leighton1969}
{Leighton}, R.~B. 1969, \apj, 156, 1, \dodoi{10.1086/149943}

\bibitem[{Lemen {et~al.}(2012)Lemen, Title, Akin, Boerner, Chou, Drake, Duncan,
  Edwards, Friedlaender, Heyman, Hurlburt, Katz, Kushner, Levay, Lindgren,
  Mathur, McFeaters, Mitchell, Rehse, Schrijver, Springer, Stern, Tarbell,
  Wuelser, Wolfson, Yanari, Bookbinder, Cheimets, Caldwell, Deluca, Gates,
  Golub, Park, Podgorski, Bush, Scherrer, Gummin, Smith, Auker, Jerram, Pool,
  Soufli, Windt, Beardsley, Clapp, Lang, \& Waltham}]{Lemen2012}
Lemen, J.~R., Title, A.~M., Akin, D.~J., {et~al.} 2012, Solar Physics, 275, 17,
  \dodoi{10.1007/s11207-011-9776-8}

\bibitem[{Lin {et~al.}(2004)Lin, Kuhn, \& Coulter}]{Lin_2004}
Lin, H., Kuhn, J.~R., \& Coulter, R. 2004, The Astrophysical Journal, 613,
  L177, \dodoi{10.1086/425217}

\bibitem[{Linker {et~al.}(2017)Linker, Caplan, Downs, Riley, Mikic, Lionello,
  Henney, Arge, Liu, Derosa, Yeates, \& Owens}]{Linker_2017}
Linker, J.~A., Caplan, R.~M., Downs, C., {et~al.} 2017, The Astrophysical
  Journal, 848, 70, \dodoi{10.3847/1538-4357/aa8a70}

\bibitem[{{Lowder} {et~al.}(2017){Lowder}, {Qiu}, \& {Leamon}}]{Lowder_2017}
{Lowder}, C., {Qiu}, J., \& {Leamon}, R. 2017, \solphys, 292, 18,
  \dodoi{10.1007/s11207-016-1041-8}

\bibitem[{Nolte {et~al.}(1976)Nolte, Krieger, Timothy, Gold, Roelof, Vaiana,
  Lazarus, Sullivan, \& McIntosh}]{Nolte1976}
Nolte, J.~T., Krieger, A.~S., Timothy, A.~F., {et~al.} 1976, Solar Physics, 46,
  303, \dodoi{10.1007/BF00149859}

\bibitem[{Obridko {et~al.}(2009)Obridko, Shelting, Livshits, \&
  Asgarov}]{Obridko2009}
Obridko, V.~N., Shelting, B.~D., Livshits, I.~M., \& Asgarov, A.~B. 2009, Solar
  Physics, 260, 191, \dodoi{10.1007/s11207-009-9435-5}

\bibitem[{{Parker}(1955)}]{Parker_1955}
{Parker}, E.~N. 1955, \apj, 122, 293, \dodoi{10.1086/146087}

\bibitem[{Priest(2014)}]{priest2014}
Priest, E. 2014, Magnetohydrodynamics of the Sun (Cambridge University Press).
\newblock \url{https://books.google.ru/books?id=BrbSAgAAQBAJ}

\bibitem[{Reiss {et~al.}(2014)Reiss, Temmer, Rotter, Hofmeister, \&
  Veronig}]{Reiss_2014}
Reiss, M., Temmer, M., Rotter, T., Hofmeister, S., \& Veronig, A. 2014, Cent.
  Eur. Astrophys. Bull., 38

\bibitem[{Robbins {et~al.}(2006)Robbins, Henney, \& Harvey}]{Robbins2006}
Robbins, S., Henney, C.~J., \& Harvey, J.~W. 2006, Solar Physics, 233, 265,
  \dodoi{10.1007/s11207-006-0064-y}

\bibitem[{Ronneberger {et~al.}(2015)Ronneberger, P.Fischer, \& Brox}]{unet}
Ronneberger, O., P.Fischer, \& Brox, T. 2015, in LNCS, Vol. 9351, Medical Image
  Computing and Computer-Assisted Intervention (MICCAI) (Springer), 234--241.
\newblock \url{http://lmb.informatik.uni-freiburg.de/Publications/2015/RFB15a}

\bibitem[{{Scherrer} {et~al.}(2012){Scherrer}, {Schou}, {Bush}, {Kosovichev},
  {Bogart}, {Hoeksema}, {Liu}, {Duvall}, {Zhao}, {Title}, {Schrijver},
  {Tarbell}, \& {Tomczyk}}]{Scherrer_2012}
{Scherrer}, P.~H., {Schou}, J., {Bush}, R.~I., {et~al.} 2012, \solphys, 275,
  207, \dodoi{10.1007/s11207-011-9834-2}

\bibitem[{Scholl \& Habbal(2008)}]{Scholl2008}
Scholl, I.~F., \& Habbal, S.~R. 2008, Solar Physics, 248, 425,
  \dodoi{10.1007/s11207-007-9075-6}

\bibitem[{Solanki {et~al.}(2006)Solanki, Inhester, \&
  Schüssler}]{Solanki_2006}
Solanki, S.~K., Inhester, B., \& Schüssler, M. 2006, Reports on Progress in
  Physics, 69, 563–668, \dodoi{10.1088/0034-4885/69/3/r02}

\bibitem[{Stenflo(2013)}]{Stenflo_2013}
Stenflo, J. 2013, Astronomy and Astrophysics Review, 21,
  \dodoi{10.1007/s00159-013-0066-3}

\bibitem[{{Tlatov} {et~al.}(2014){Tlatov}, {Tavastsherna}, \&
  {Vasil'eva}}]{Tlatov_2014}
{Tlatov}, A., {Tavastsherna}, K., \& {Vasil'eva}, V. 2014, \solphys, 289, 1349,
  \dodoi{10.1007/s11207-013-0387-4}

\bibitem[{Toma(2010)}]{Toma_2011}
Toma, G. 2010, Solar Physics - SOL PHYS, 274, \dodoi{10.1007/s11207-010-9677-2}

\bibitem[{{Toma} \& {Arge}(2005)}]{Toma_2005}
{Toma}, G.~D., \& {Arge}, C.~N. 2005, in Astronomical Society of the Pacific
  Conference Series, Vol. 346, Large-scale Structures and their Role in Solar
  Activity, ed. K.~{Sankarasubramanian}, M.~{Penn}, \& A.~{Pevtsov}, 251

\bibitem[{{Verbeeck, C.} {et~al.}(2014){Verbeeck, C.}, {Delouille, V.},
  {Mampaey, B.}, \& {De Visscher, R.}}]{Verbeeck_2014}
{Verbeeck, C.}, {Delouille, V.}, {Mampaey, B.}, \& {De Visscher, R.} 2014,
  A\&A, 561, A29, \dodoi{10.1051/0004-6361/201321243}

\bibitem[{{Vr{\v{s}}nak} {et~al.}(2007){Vr{\v{s}}nak}, {Temmer}, \&
  {Veronig}}]{Vrsnak_2007}
{Vr{\v{s}}nak}, B., {Temmer}, M., \& {Veronig}, A.~M. 2007, \solphys, 240, 315,
  \dodoi{10.1007/s11207-007-0285-8}

\bibitem[{{Wiegelmann} {et~al.}(2017){Wiegelmann}, {Petrie}, \&
  {Riley}}]{Wiegelmann_2017}
{Wiegelmann}, T., {Petrie}, G. J.~D., \& {Riley}, P. 2017, \ssr, 210, 249,
  \dodoi{10.1007/s11214-015-0178-3}

\bibitem[{Wiegelmann {et~al.}(2014)Wiegelmann, Thalmann, \&
  Solanki}]{Wiegelmann_2014}
Wiegelmann, T., Thalmann, J.~K., \& Solanki, S.~K. 2014, The Astronomy and
  Astrophysics Review, 22, \dodoi{10.1007/s00159-014-0078-7}

\end{thebibliography}
\bibliographystyle{aasjournal}



\end{document}